\documentclass[fleqn,usenatbib]{mnras}

\usepackage{newtxtext,newtxmath,amsmath}
\usepackage[T1]{fontenc}
\usepackage{ae,aecompl}
\usepackage{xcolor}
\usepackage{hyperref} 
\usepackage{graphicx}	% Including figure files

\newcommand{\ms}{M$_{\odot}$}
\newcommand{\zs}{Z$_{\odot}$}
\newcommand{\afe}{[$\alpha$/Fe]}
\newcommand{\feh}{[Fe/H]}
\title[Evolution and chemical properties of the Galactic discs]
{On the origin of the Galactic thin and thick discs, their abundance gradients and the diagnostic potential of their abundance ratios }
\author[Prantzos et al.]{
Nikos Prantzos$^{1}$ \thanks{E-mail: prantzos@iap.fr },
Carlos Abia$^{2}$,
Tianxiang Chen$^{1}$,
Patrick de Laverny$^{3}$,
Alejandra Recio-Blanco$^{3}$,
\newauthor
E. Athanassoula$^4$,
Lorenzo Roberti$^{5,6}$,
Diego Vescovi$^{7,8}$,
Marco Limongi$^9$,
Alessandro Chieffi$^9$,
Sergio Cristallo$^{7,8}$
\\
1. Institut d'Astrophysique de Paris, CNRS and Sorbonne Universit\'e, 75014 Paris, France 
\\
2. Dpto. F\'\i sica Te\'orica y del Cosmos, Universidad de Granada, 18071 Granada, Spain
\\
3. Universit\'e C\^ote d’Azur, Observatoire de la C\^ote d’Azur, CNRS, Laboratoire Lagrange, France
\\
4. Aix Marseille Univ, CNRS, CNES, LAM, Marseille, France
\\
5. Konkoly Observatory, Research Centre for Astronomy and  Earth Sciences, E\"otv\"os Lor\'and Research Network (ELKH), H-1121 Budapest, Hungary
\\
6. CSFK, MTA Centre of Excellence, Budapest,  H-1121, Hungary
\\
7. Istituto Nazionale di Astrofisica - Osservatorio Astronomico d’Abruzzo, Via Maggini snc, I-64100, Teramo, Italy
\\
8. INFN -  Sezione di Perugia, via A. Pascoli, Perugia, Italy
\\
9. Istituto Nazionale di Astrofisica—Osservatorio Astronomico di Roma, Frascati, Italy 
\\
}
\date{Accepted XXX. Received YYY; in original form ZZZ}

\pubyear{2023}

\begin{document}
\label{firstpage}
\pagerange{\pageref{firstpage}--\pageref{lastpage}}
\maketitle

\begin{abstract}
 Using a semi-analytical model of the evolution of the Milky Way, we show how secular evolution can create distinct overdensities  in the phase space of various properties (e.g. age vs metallicity or abundance ratios vs age) corresponding to the thin and  thick discs. In particular, we show how  key properties of the Solar vicinity can be obtained by secular evolution, with no need for external or special events, like galaxy mergers or paucity in star formation. This concerns the long established double-branch behaviour of [alpha/Fe] vs metallicity and  the recently found non-monotonic evolution of the stellar abundance gradient,      evaluated at the birth radii of stars.  We extend the discussion to other abundance ratios and we suggest a classification scheme, based on the nature of the corresponding yields (primary vs secondary or odd elements) and on the lifetimes of their  sources (short-lived vs long-lived ones). The latter property is critical in determining the single- or double- branch behavior of an elementary abundance ratio in the Solar neighborhood. We underline the high diagnostic potential of this finding, which can help to separate clearly elements with sources evolving on different timescales and help determining  the site of e.g. the r-process(es). We define the "abundance distance" between the thin and thick disc sequences as an important element for such a separation. We also show how the inside-out evolution of the Milky Way disc leads rather to a single-branch behavior in other disc regions. 

\end{abstract}


%%%%%%%%%%%%%%%%%%%%%%%%%%%%%%%%%%%%%%%%%%%%%%%%%%

%%%%%%%%%%%%%%%%% BODY OF PAPER %%%%%%%%%%%%%%%%%%

\section{Introduction}
\label{sec:Intro}

Nowadays there is an exponential increase in the amount of information regarding the stellar populations of our Galaxy. This is due to several large scale surveys that started during the last decade, like RAVE \citep{rav20}, APOGEE \citep{Abdurro2022APOGEE}, GALAH \citep{Buder2021}, LAMOST \citep{lam22}, Gaia-ESO \citep{gae22} and, last but not least, the results of ESA's Gaia mission \citep{gaia22}. Those surveys provided information for hundreds of thousands of stars in the 6D space of stellar positions and velocities and in the multi-dimensional space of elemental abundances, mostly for giants but also for dwarf stars. The determination of another key stellar property, namely stellar ages, still suffers from considerable uncertainties, but the situation is expected to improve soon due to progress in asteroseismology \citep[as illustrated in e.g.][]{Miglio2021}.

Those observations shed new light on our understanding of the structure of the Milky Way  \citep[MW, e.g.][and references therein]{Bland2016}. In particular, they have revealed structures in the kinematic and chemical  phase space, attributed to debris from old collisions/mergers of various galactic sub-systems with the Galaxy; despite some overlap in their properties, such structures provide important hints on the past history of our Galaxy and the links between its various components \citep[see][for a recent overview]{Helmi2020}.

Among the main baryonic components of the MW (bulge, bar,  stellar halo, discs), the thick disc remains the most enigmatic. Since its identification as a distinct entity through observations of its vertical density profile \citep{Gilmore1983}, its various properties have been extensively studied, both locally and across the whole Galaxy. It is now clear that it differs from the thin disc in all aspects of its stellar population: spatial, kinematic, metallicity, elemental abundance ratios and stellar ages. The thick disc is, on average, older than the thin disc (age $\sim$10 Gy compared to $\sim4-5$ Gy), less metallic, with a mean [Fe/H]\footnote{We adopt here the usual notation [X/H]$=$ log (X/H)$_\star-$ log (X/H)$_\odot$, where (X/H)$_\star$ is the abundance by number of the element X in the corresponding star.}  $\sim-0.6$ dex compared to $\sim 0$ dex, it has  higher \afe \ at a given metallicity, larger velocity dispersion and larger scale-height.
However, in each one of those properties there is an overlap with the thin disc  and it is difficult to identify unambiguously the  stars of the two discs. As discussed in \cite{Kawata2016} there is no clear definition  for the thin and thick discs, even in the case of the MW where ample information is available; the answer depends a lot on how the discs are defined (chemically, kinematically, geometrically or by age). As a result of this ambiguity, there is still considerable uncertainty on the contribution of the thin and thick discs to the  MW mass budget \citep{Anguiano2020,Everall2022,Vieira2022}.

An early attempt to model the formation of thick and thin discs of the MW was made by \cite{Burkert1992} who found that it naturally results  as a consequence of the gravitational settling and self-regulated chemical and dynamical evolution of an initially hot, gaseous protodisc (top-down formation with high early star formation rate). On the other hand, \cite{Quinn1993}, studying N-body simulations of  halo-disc systems colliding with satellite galaxies found that the kinematical heating induced by such mergers may lead to the formation of thick discs with properties (scale-height, vertical velocity distribution, asymmetric drift) similar to those observed in the MW. 
Using a semi-analytical, multi-zone model \cite{Chiappini1997} assumed two periods of star formation, a short and early one (thick disc with high SFR) separated by $\sim$2 Gy hiatus from a longer one  (thin disc with low SFR); their scheme produced a small loop in the [O/Fe] vs \feh \ relation at 4 kpc, but did not modify that relation in the outer disc or the Solar neighborhood. \cite{Kroupa2002b} suggested that the high early SFR  (induced by external mergers or perturbations) produced more massive star clusters and  high mass stars than the later, low SFR period; the subsequent relaxation of and gas expulsion from those clusters left a characteristic kinematic signature, namely the observed high velocity dispersion of the stars of the thick disc.

%The action of the bar can mix radially not only gas but also stars, and the effects of stellar radial motions on the abundance profiles have been studied to some extent with N-body+SPH codes by \cite{Friedli1993} and \cite{Friedli1994}. Observations in the 90s revealed that the  MW does have a bar \citep{Blitz1991}, but its origin, size and age are not well known yet; as a result, its impact on the evolution of the MW is difficult to evaluate quantitatively.

 \cite{SellwoodBinney2002}  showed  that,  in the presence of recurring transient spirals, stars in a galactic disc could 
undergo important radial displacements. Stars 
found at corotation with a spiral arm may  be scattered to different galactocentric radii (inwards or outwards),  a process called {\it churning} that preserves
overall angular momentum distribution and does not contribute to the radial heating of the stellar disc, in contrast to simple epicyclic motion ({\it blurring}) which does heat the disc radially.
\cite{Schon2009} introduced a parametrised prescription of radial migration (distinguishing epicyclic motions 
from migration  due to transient spirals) in a semi-analytical chemical evolution code.
They suggested that radial mixing could also explain the formation of the Galaxy's thick disc, by bringing   a kinematically "hot" stellar population from the inner disc to the Solar neighbourhood. 
That possibility was subsequently investigated
with N-body models, but controversial results have been  obtained. While \cite{Roskar2008} and \cite{Loebman2011} found that secular processes (i.e. radial migration) are sufficient to explain the kinematic properties of the local thick disc, \cite{Minchev2012b} found this mechanism insufficient  and suggested that an external agent, like early mergers, is required for that \citep[see also][]{Minchev2016}. 

Despite the large number of studies over the years, it still remains unclear the role played  by  
the various processes invoked to explain the morphological, kinematic and chemical properties of the thick and  thin discs of the MW: 
a clumpy and turbulent early galaxy \cite[e.g.][]{Agertz2009,Bournaud2009,Agertz2021_VINTER_I,Renaud2021_VINTER_II,Beraldo2021}; radial stellar migration, accompanied or not by gas migration \citep[e.g.][]{Schon2009,Loebman2011,Kubryk2015a,Sharma2021b}; internal "heating of the disc" \citep{Kroupa2002b}; various types of mergers, "dry" or "wet", minor or major ones,
 \citep[e.g][]{Villalobos2008,Wilson2011,Navarro2011,Bekki2011,Brook2012,Forbes2012,Bird2013,Athanassoula2016, Grand2018,Grand2020,Belokurov2022}); major episodes of infall and/or star formation \citep{Noguchi2018,Buck2020,Vincenzo2020,Khoperskov2021,Conroy2022}  or various combinations of the above. For instance, 
\citet[][and in preparation]{Athanassoula2016} used high-resolution hydrodynamic N-body simulations to follow intermediate, or even major, gas-rich mergers and the corresponding formation and evolution of the thick and thin discs; they found two periods of star formation: an early one, intense and short, and a late one,  extended in time but of low amplitude.

 On the other hand,  using mono-abundance populations (i.e. defined in the plane of [O/Fe] vs. [Fe/H]), \cite{Bovy2012} concluded that the thick disc is not really a distinct component of the MW, \citep[see also][for a review]{Rix2013}. The study of \cite{Park2021} with high-resolution cosmological simulations of 19 large disc galaxies finds that thick and thins discs are commonly formed as
"two parts of a single continuous disc component that evolves
with time as a result of the continued star formation of thin disc
stars and disc heating". However, in a recent  data analysis  of Gaia eDR3 and Lamost (DR7),  \cite{XiangRix2022} find distinctive signatures in the MW age-metallicity relation, suggesting that the major merger  with Gaia-Enceladus played a significant role in the thick disc formation. It is not clear, however, whether such major events contributed to enhance or rather to inhibit star formation in those early epochs, thus initiating or terminating the thick disc formation \cite[e.g.][]{Conroy2022,Ciuca2022}. Furthermore, it is not clear whether the formation of the thick disc preceded the one of the thin disc, as commonly assumed, or whether the two were co-eval as suggested in recent studies \citep{Beraldo2021,Gent2022}.

The double-branch behaviour of \afe \ ratio in the local disc \citep{Fuhrmann1998,Bensby2005,Bensby2014} attracted a lot of attention, being one of the clearest signatures  helping to differenciate observationally the thick from the thin disc \citep[e.g.][]{Adibekyan2012}. This dichotomy has been investigated with both semi-analytical models and chemodynamical ones (with a small fraction of the latter class being referenced in the previous paragraphs).
Semi-analytical models can be broadly classified in two categories: 

a) one class of models  invokes two periods of star formation in the MW, one corresponding to the thick and one to the thin disc. Those periods are characterised by distinct star formation histories, a shorter early one with intense star formation and a longer late one with a lower star formation, separated by a pause ("hiatus") of several Gy \citep{Noguchi2018,Grisoni2017,Lian2020MNRAS,Spitoni2021}. In those models 
%the stars of both discs are formed in the same place (no radial stellar motions between the model zones) and  
the double-branch of  \afe \ vs \feh \ is obtained by assuming that at the end of the first phase a strong, metal-poor  infall dilutes the gas abundances
to considerable amounts (by a factor of 3 at least) and the subsequent star formation creates another \afe \ vs \feh\- sequence, more or less parallel to the first and starting from lower metallicities and lower \afe \ ratios.

b) the second class is composed of "hybrid" models: multi-zone semi-analytical models of chemical evolution with radial migration introduced in various ways \citep{Schon2009,Kubryk2015a,Johnson2021,Sharma2021b,Weinberg2019,Weinberg2022}; in most cases the dynamics are not treated self-consistently with the chemical evolution and various approximations are made. The two branches of \afe \ vs \feh \ in the Solar neighborhood result then from the presence of stars  formed not only locally but also in the inner Galaxy, where intense star formation takes place early on.
In contrast to the previous case, this is obtained by secular evolution. However, to explain the different dynamics of the thick disc still some different physical processes in the early Galaxy need to be invoked, like e.g. an early highly turbulent or bursty phase, or perturbations by mergers \citep[as in][]{Minchev2013,Minchev2018}. 

The evolutionary scheme of category (a) seems to be supported by the analysis  of \cite{Vincenzo2021}, based on  the abundances of a sample of red giants from APOGEE DR16. They conclude that
the observed bimodality of \afe \ vs \feh \ probably requires one or more sharp transitions in the disc's gas accretion, star formation, or outflow history in addition to radial mixing of stellar populations. Similarly,  \cite{Lu2022b} analyze $\sim$80000 subgiant stars in LAMOST and find a non-monotonic behaviour of the \feh \  gradient with age around 8-11 Gy ago. They associate it to the last major merger (Gaia Sausage/Enceladus event) and they  claim that this transition plays a major role in shaping the [$\alpha$/Fe] vs [Fe/H] dichotomy.

The evolutionary scheme of category (b) is supported by the conclusion of \cite{Hayden2017}, based on the orbital and chemical properties of $\sim$500 stars from the AMBRE survey, 
that "{\it the high-\afe \ and low-\afe \ sequences are most likely a reflection of the chemical enrichment history of the inner and
outer disc populations, respectively; radial mixing causes both populations to be observed in situ at the Solar position}". 

In this work we present a model in line with \cite{Hayden2017}, an updated version of the
one presented in \cite{Kubryk2015a} (Sec. \ref{sec:Model}). We present in some detail its main properties in Sec. \ref{sec:Model_discs}, showing how secular evolution can create distinct overdensities, corresponding to the thin and  thick discs, in the phase space of various 
properties (e.g. age vs metallicity or abundance ratios vs age). We also show how the non-monotonic behaviour of the stellar [Fe/H] gradient of stars -{evaluated at their bith place}- with age  by \cite{Lu2022b} can be explained by secular evolution and  we identify the conditions for that. In Sec.  \ref{sec:alpha_Fe} we focus on the \afe \ vs \feh \ diagram obtained for the Solar neighborhood and we show how the double-branch behaviour can also be obtained by secular evolution. Then in Sec. \ref{sec:OtherXY} we extend the discussion to other abundance ratios and we suggest that in all studies of that kind, the "distance" in chemical space between thin and thick discs should be quantified in order to clearly identify a single- or double-branch behaviour for a given element. We further suggest a classification scheme of the chemical elements, based on the nature of the corresponding yields (primary vs secondary or odd elements) and on the lifetimes of their sources (short-lived vs long-lived ones). We argue that the latter property is critical in determining the single- or double- branch behaviour of an elementary abundance ratio in the Solar neighborhood. We underline the high diagnostic potential of this finding, which can help to separate clearly elements with sources evolving on different timescales; it would be interesting, in that respect to have a precise evaluation of the abundance distance between the thin
and thick disc sequences for the r-elements, which have been suggested to originate either in short-lived sources (collapsars) or long-lived ones (neutron star mergers, e.g. \cite{Wanajo2021} and references therein). We also show how the inside-out evolution of the MW disc leads rather to a single-branch behaviour in other disc regions, dominated either by the "high" \afe \ branch (inner disc) or the "low" \afe\- branch (outer disc). In Sec \ref{subsec:OtherXY-discusssion} we discuss other recent studies and we summarize our results in Sec. \ref{sec:Summary}.

\section{The model}
\label{sec:Model}

\subsection{Model}
\label{subsec:Model}

We summarize here the main features of the \cite{Kubryk2015a} model and the new ingredients adopted in this work.

\subsubsection{Stars and gas}
\label{subsub:StarsGas}
\begin{figure}
	\includegraphics[width=0.49\textwidth]{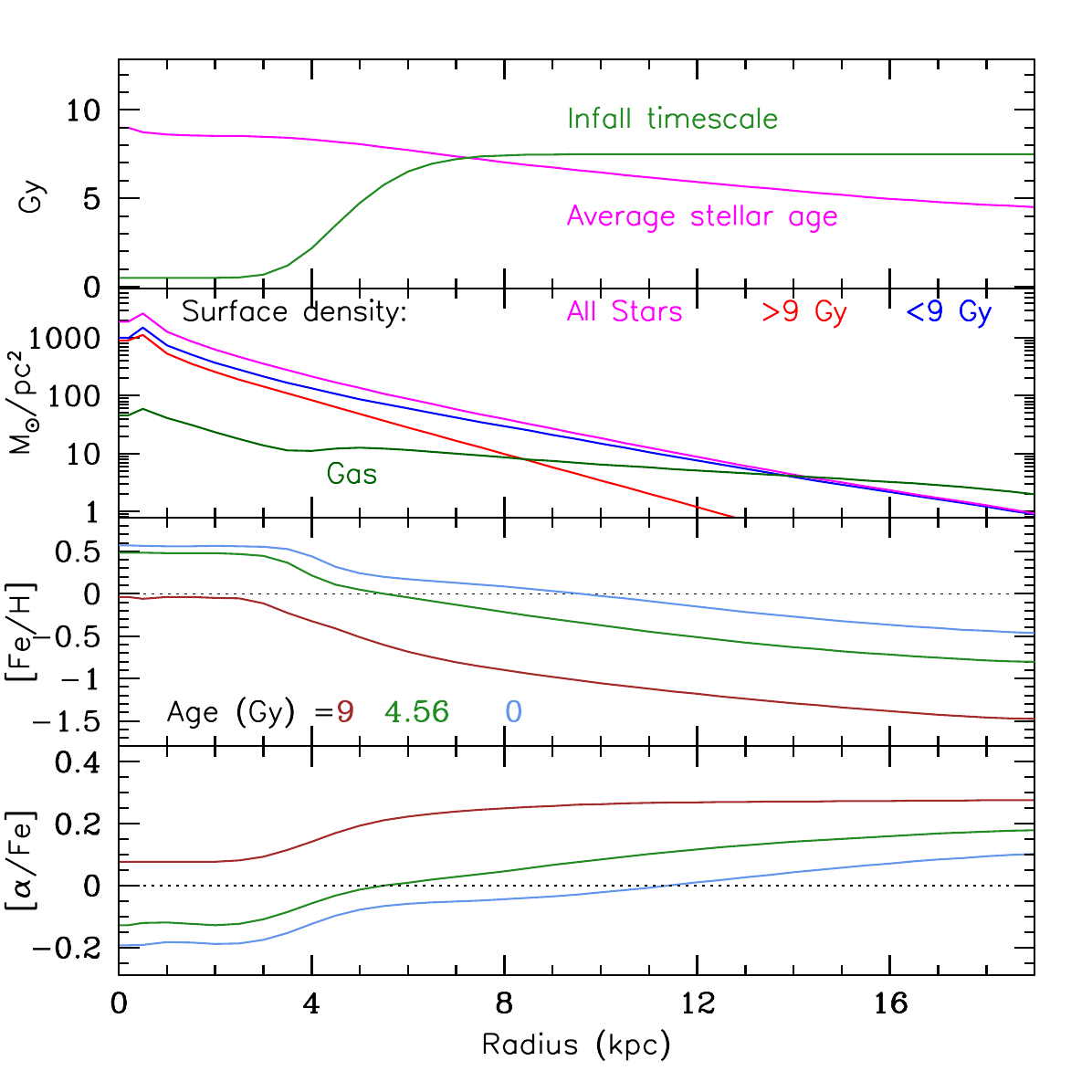}
    \caption{ {\it Top panel:}  Adopted infall time-scale (green) and resulting average stellar age (solid magenta)   at the end of the simulation. {\it Second panel:} Final surface density mass profile of gas (green), "young" stars (age $<9 $ Gy, blue), "old" stars (age$>9$ Gy, red) and all stars (magenta). {\it Third panel}: Fe abundance profiles in the gas, 9 Gy ago (red), at Sun's formation (4.56 Gy ago, green) and today (blue). {\it Bottom:} Gaseous [$\alpha$/Fe] profiles, colour coded as in previous panel.
    }
    \label{fig:f1_discs_gen}
\end{figure}

The Galactic disc is gradually built up by infall of primordial gas in the potential well of a dark matter halo with mass of $10^{12}$ \ms, the evolution of which - mass growing with time - is obtained from
numerical simulations \citep[from][]{Li2007}. 
The one-dimensional (1D) disc is built "inside-out" through infall of gas of primordial composition. The  infall time-scales are fairly short (0.5 Gy)  in the inner regions and increase progressively outwards,  reaching 7.5 Gy at 8 kpc and staying nearly constant after that (see top panel in Fig. \ref{fig:f1_discs_gen}).
The star formation rate depends on the local surface
density of molecular gas $\Sigma_{H_2}$, the latter being calculated from the total gas through the prescription of \cite{Blitz2006}.

The disc is split in concentric annuli of radial width $\Delta R=0.5$ kpc, interacting through radial motions of their stars.
Stars move radially from their birthplace due to epicyclic motions around their guiding radius (blurring) and to changes of their guiding radius (churning), as explained below. 
The innovative aspect of the model is  that it accounts for the fact that radial migration moves around not only "passive tracers" of chemical evolution (i.e. long-lived low-mass stars, keeping in their photospheres the chemical composition of the gas at the time and place of their birth), but
also "active agents" of chemical evolution, i. e., long-lived nucleosynthesis sources
(such as SNIa producing Fe and low mass stars producing s-process elements), which may release their products away from their birth place. The probabilistic formalism for radial migration takes into account the finite lifetime of stars \citep[see][]{Kubryk2015a}.  In practice,  stars more massive than 4 \ms (with lifetimes $<$0.3 Gy) do not live long enough to participate in such motions and die in their birth zone (here taken to be 0.5 kpc wide). This also applies to giant molecular clouds, their lifetime being comparable to those of massive stars, of the order of $10-30$ Ma \citep{Jeffreson2021}.

%The model takes into account the radial flows of gas driven by a bar formed 6 Gy ago which pushes gas inwards and outwards of the corotation. 

In the \cite{Kubryk2015a,Kubryk2015b} model, the thick disc is assumed to be the oldest part of the Galactic disc, older than $\sim 9 $ Gy  \citep[see][]{Binney2014}, while the thin disc is younger than 9 Gy.  In this work, the thin and thick discs appear naturally as overdensities in the phase space of various  properties of stars in the Solar neighborhood (age, metallicity, abundance ratios) as we discuss in the next sections, and they do not result from a pause in star formation. Instead, they are produced in a continuous inside-out formation process, as suggested by the recent simulations of \cite{Park2021}. These simulations have unprecedentedly high spatial resolution
(aimed to reach $\Delta$x$\sim$ 34 pc at redshift $z= 0$), making them ideal to investigate the detailed structures
of galaxies. \cite{Park2021} find that spatially-defined thin and
thick discs are not entirely distinct components in terms of formation process, but  rather two parts of a single continuous disc component that evolves with time as a result of the continued star formation of thin-disc stars
and disc heating.

The radial displacements of stars due to blurring and churning are calculated statistically, with prescriptions adopted on the  basis of N-body SPH simulations \citep{Kubryk2013,Kubryk2015a}.
 Blurring is calculated by using the stellar radial velocity dispersion, which is assumed to evolve with time as $\sigma_r \propto \tau^{\beta}$, but in contrast to \cite{Kubryk2015a} where $\beta=0.33$, here we adopt 
$\beta=0.25$, based on analysis of numerical simulations of the heating of MW-type discs by \cite{Aumer2016} and of MW observations from GALAH, LAMOST, APOGEE, the NASA Kepler and K2 missions, and Gaia DR2 from \cite{Sharma2021}. We assume that the old disc (age >9 Gy today) inherited a  large radial velocity dispersion of $\sigma_r(R)=50 {\rm e^{-(R-8)/11)}}$ km/s, where R is the radius in kpc and 11 kpc is the characteristic scale of the decline of radial dispersion of the thick disc (see  Bland et al (2016), and references therein\footnote{See \cite{Mackereth2019} for a more detailed analysis of velocity dispersions of the MW.}). Such a high dispersion can be acquired either during an initial, highly turbulent phase of the gas  \citep[e.g.][]{Bournaud2009,Brucy2020}, through a major merger around 8-9 Gy ago, like  Gaia-Enceladus  \citep[][and references therein]{Helmi2020}, or through some  combination  of various physical processes. 

Indeed, the recent analysis  by \cite{Yan2019} of a sample of 307,246
A, F, G, K-type giant stars from the LAMOST spectroscopic survey and Gaia DR2 survey concludes that
the formation of the thick disc could be affected by more than one process: the
accretion model seems to have played a prominent role, but other formation mechanisms, such as  radial migration
or disc heating  could also have a contribution. Also, \cite{Beraldo2021}, analyzing the kinematics and orbits of 23795 turnoff and giant stars with
6-D phase-space coordinates from Gaia-DR2 find that in the Solar neighbourhood, about half of the old thin disc stars
can be classified as migrators, while for the thick disc this migrating fraction could be as high as $\sim$1/3.

\subsubsection{Nucleosynthesis sources}
\label{subsub:Chemistry}

The \cite{Kubryk2015a,Kubryk2015b} version of the model uses  the metallicity-dependent yields of \cite{Nomoto2013} for massive stars with no mass loss, and \cite{Karakas2010} for low and intermediate masses (LIMS), which are calculated up to a metallicity of Z$=2$ \zs, and are appropriate for the study of the inner Galactic disc. However, in \cite{Karakas2010} the s-process is not considered.  \cite{Prantzos2018,Prantzos2020}   adopted for LIMS the yields of \cite{Cr15} which include the s-process and are calculated up to Z$=2\times 10^{-2}\sim1.45$ \zs,  and for massive stars those of  \cite{Limongi2018}, which include the effect of mass loss and  stellar rotation but go only up to \zs. Here we use the same yields augmented by a new set of models at Z$=3 \times 10^{-2}\sim2$ \zs, calculated for this study by the same authors and available in the corresponding databases\footnote{
\href{http://fruity.oa-abruzzo.inaf.it} {\textcolor{blue}{{http://fruity.oa-abruzzo.inaf.it}}}
and \href{http://orfeo.iaps.inaf.it/index.html}{\textcolor{blue}{{http://orfeo.iaps.inaf.it/index.html}}} for LIMS and
massive stars; respectively.}; indeed,  the study of the inner disc, which may reach metallicities as high as 2-3 \zs,  requires the use of model yields from super-Solar metallicity stars. Also, in contrast to all previous studies of radial migration plus chemical evolution in the Galactic disc, this is the first study using stellar yields from massive stars dependent on rotational velocity. The IDROV (Initial Distribution of ROtational Velocities) is presented in \citet{Prantzos2018} for a 1-zone model, where it was "calibrated" as function of metallicity in order to reproduce key observables and  it is also adopted here. We note that a comparative study of \cite{Philcox2018} found that the yields and IDROV adopted in \cite{Prantzos2018} reproduce best the proto-solar abundances -  compared to other sets of yields - for a large number of elements.

The initial mass function (IMF) of \cite{Kroupa2002} with a slope 1.3 for the high masses, is considered.  A phenomenological  rate of SNIa is adopted,  based on observations of extragalactic SNIa \cite{Maoz2017} while their  metallicity-dependent yields are from \cite{Iwamoto1999}. 
In  order  to calculate the rate of ejecta (both for stars and SNIa) as a function of time, the    formalism of single particle population is used  because it can account for  the radial displacements of nucleosynthesis sources and in particular of SNIa, as discussed in Appendix C in \cite{Kubryk2015a}.

\section{Galactic disc properties of the model}
\label{sec:Model_discs}

\subsection{Overall properties}
\label{sub:Model_StarDistr}

Some key features of the model, relevant to our discussion, are displayed in Fig. \ref{fig:f1_discs_gen}. The inner regions (here taken to be as those with Galactocentric distance R$_{\rm G}<4$ kpc) are formed quite rapidly, since the adopted characteristic timescale of gas infall is of the order of $\sim 1$ Gy (upper panel), while intermediate regions ($4<$ R$_{\rm G}<10$ kpc) are formed within  a few  Gy and the outer regions after several Gy. As a result, the average age of stars today is $\sim 9$ Gy for the inner disc, 6-8 Gy for the intermediate disc and $<5$ Gy for the outer disc. 

The final stellar surface density profiles, at T=12 Gy (second from top panel in Fig. \ref{fig:f1_discs_gen}) are quasi-exponential, with characteristic scalelengths
of $R_{T}\sim1.9$ kpc for the old disc (age$>9$ Gy),  $\sim 2.8$ kpc for the young disc ($<9$ Gy) and 2.6 kpc for the total disc. These values should be compared to \cite{Bland2016}, who suggest $R_{T}$=2.0$\pm$0.2 kpc for the thick disc and  $R_{T}$=2.6$\pm$0.5 kpc for the thin disc, which are, however, defined chemically in their case. The resulting present day gas surface density profile is essentially flat, going through a broad maximum at  R$_{\rm G}<5$ kpc, and slowly declining outwards, again in fair agreement with observations \citep[see e.g. Appendix A in][]{Kubryk2015a}.

The Fe abundance profiles in the gas ($3^{rd}$ panel from the top in Fig. \ref {fig:f1_discs_gen}) are displayed for ages 0 (today), 4.56 Gy (Sun's formation) and 9 Gy. They are  exponentially decreasing with galactocentric distance at all ages. The abundance gradient is getting smaller (in absolute value)  with time, as a consequence of the adopted inside-out star formation scheme. The present day gradient is d[Fe/H]/dR $\sim -0.06$ dex/kpc in the range 4 $< \rm R <12$ kpc, again in fair agreement with observed values \citep[e.g.][and references therein]{Minniti2020,Spina2022}. It is approximately zero in the innermost regions, where the gas fraction is extremely small and chemical evolution has reached a steady state: metal-poor gas returned from old low-mass stars dilutes the fresh metal content released by the few young massive stars. We note the flat metallicity profile in the inner 3-4 kpc of the disc at early times (the 9 Gy curve) due to the short infall timescale  assumed for those regions and the resulting uniformly high SFR; this is in agreement with the recent finding of \cite{XiangRix2022} that the star forming gas in the old disc of the MW  had small metallicity variations (less than  0.2 dex), otherwise the resulting [Fe/H] - age relation would display larger scatter than what they infer from their data analysis of Gaia eDR3 and
LAMOST DR7 surveys.

\begin{figure}
	\includegraphics[width=0.49\textwidth]{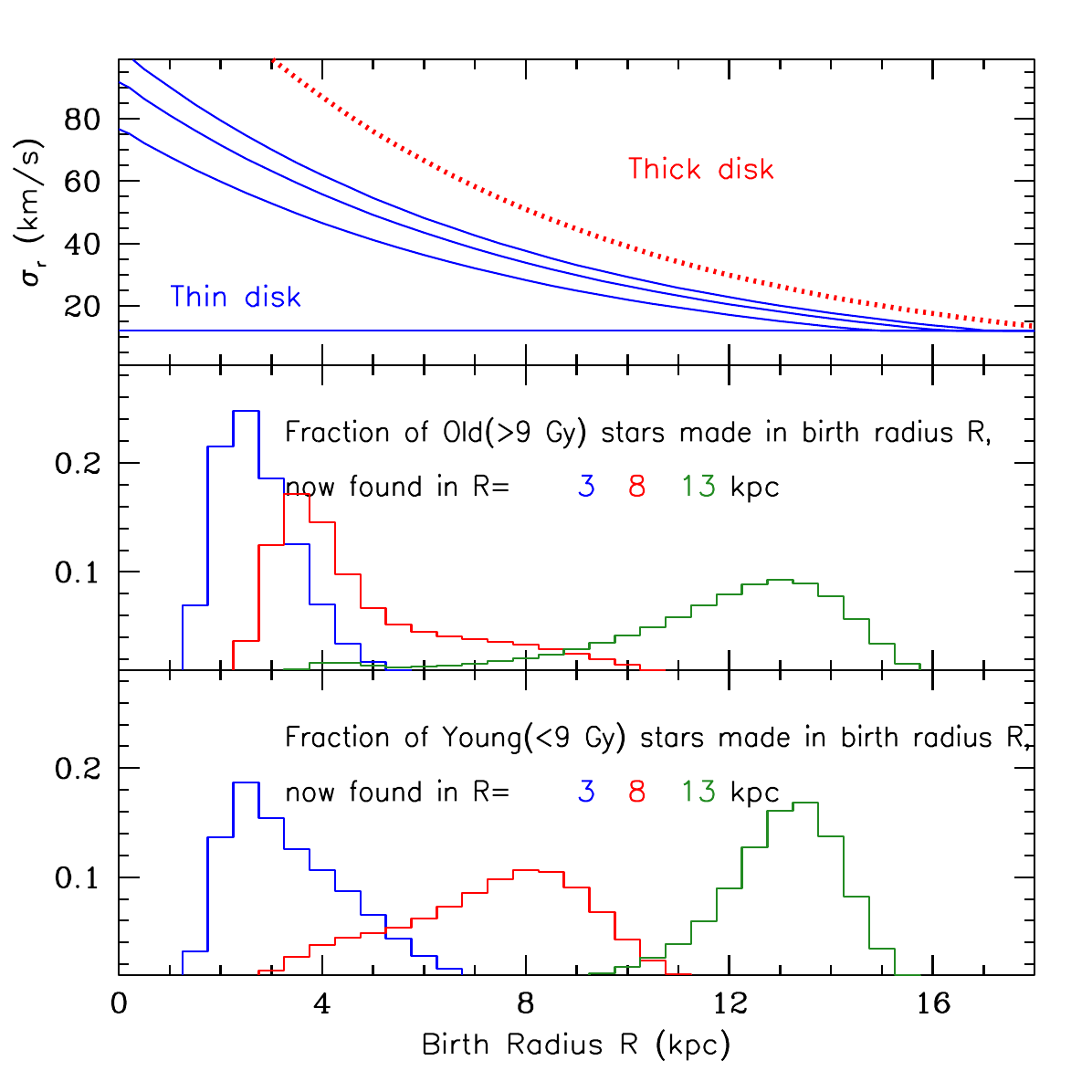}
    \caption{ {\it Top :} Radial velocity dispersion $\sigma_r$ of stellar populations in the thin disc (blue curves), after 0, 3, 6, and 9  Gy (from bottom to top); the lowest curve corresponds to the velocity dispersion of the gas. The stars of the thick disc are assumed to be formed in the first 3 Gy with a velocity dispersion indicated by the red dotted curve (see text).  {\it Middle:} Fraction of "old" stars born in radius $R$ now found in three characteristic radii R$_G =3, 8$ and 13 kpc (blue, red and green histograms, respectively). {\it Bottom:} Same as in previous panel, for "young" stars (age $<9$ Gy).
    }
    \label{fig:f2_discs_gen}
\end{figure}

Finally,  the bottom panel of Fig. \ref{fig:f1_discs_gen} displays the [$\alpha$/Fe] profiles, decreasing inwards, again as a consequence of inside-out star formation: SNIa, which are the main Fe source,  enrich the inner regions with Fe for a larger fraction of the time than the outer ones, which form their stars later. Thus, their time-integrated Fe ejecta per unit mass of gas processed is higher in the inner Galaxy and the [$\alpha$/Fe] ratio is lower there.

Fig. \ref{fig:f2_discs_gen} displays some of the  properties of the model and their consequences for the final stellar distributions as function of Galactocentric radius. In the top panel is displayed the evolution of the adopted profiles of radial dispersion of stars as function of time. For the first 3 Gy of galactic evolution (the thick disc), stars are assumed to be born 
with a time-independent radial velocity dispersion $\sigma_r(R)=50 {\rm e^{-(R-8)/11)}}$ km/s  (see Sec. \ref{subsub:StarsGas}), here shown by a red dashed curve. For the subsequent 9 Gy (the thin disc), stars are assumed to inherit at their birth the velocity dispersion of the "settled" gas  (flat profile at $\sigma_r(R)=10$ km/s,  corresponding to the gas dispersion today), then $\sigma_r$ increases proportionally to $\tau^{0.25}$, as shown here for times $\tau=$3, 6, and 9  Gy after stellar birth; the latter curve corresponds to the present-day $\sigma_r(R)$ profile of the MW disc.

\begin{figure}
	\includegraphics[width=0.49\textwidth]{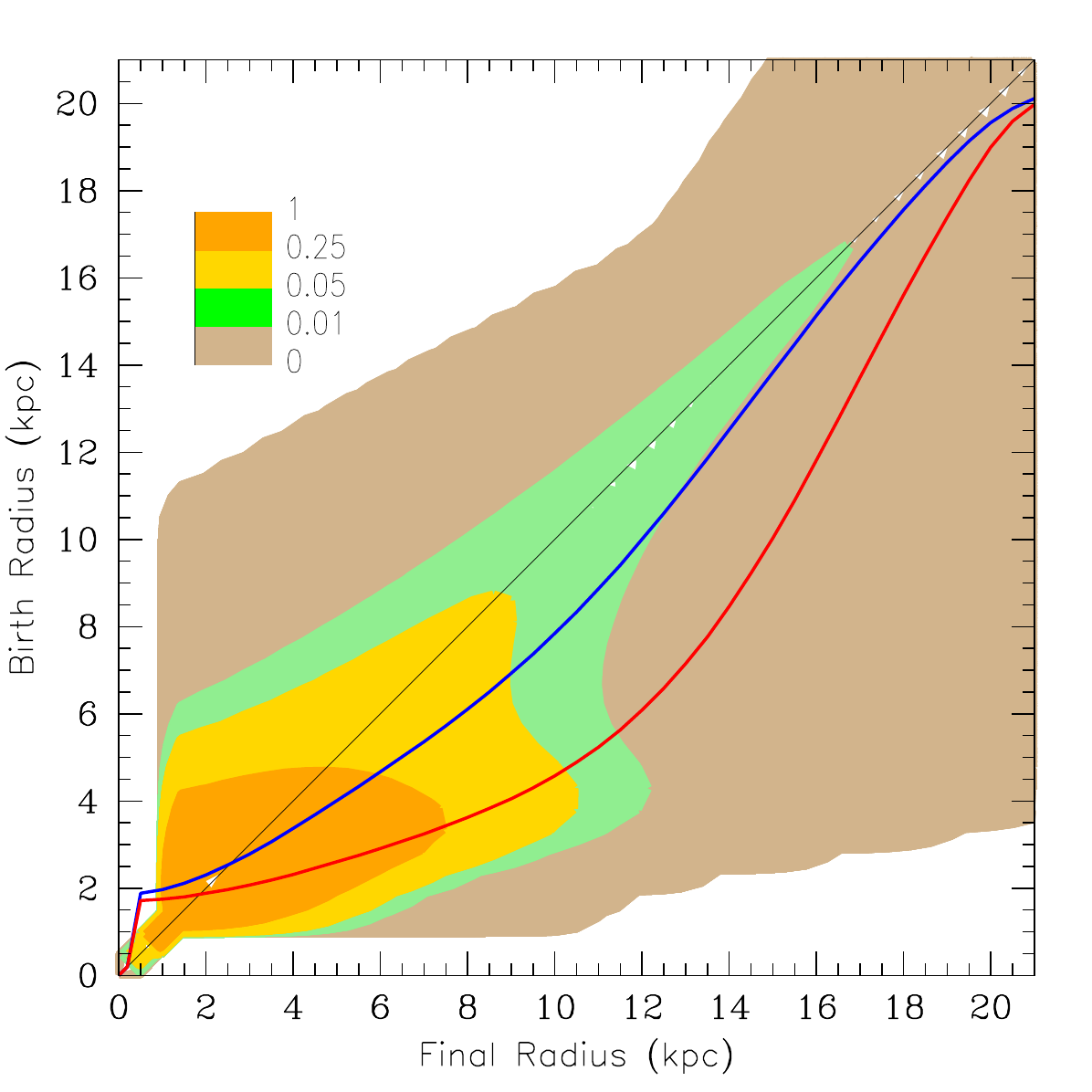} %{a00.png}
    \caption{Isocontours of star numbers (N$_{\rm max}$=1) in the plane of birth radius vs final radius at the end of the simulation (T=12 Gy). The average birth radius of stars at each final radius is indicated by the blue curve for the young disc (age$<9$ Gy) and by the red curve for the old disc (age$>9$ Gy). In the insert box are depicted the isocontour levels, also adopted in all subsequent figures (unless otherwise stated). }
    \label{fig:f2_Rb_vs_Rf}
\end{figure}

The two other panels of Fig. \ref{fig:f2_discs_gen} indicate the fraction of stars present today in the annuli centered at Galactocentric radii R$=3$,  8 and 13 kpc (blue, red and green histograms, respectively) as a function of the corresponding birth radii. For the "old" stars ($>9$ Gy), here qualified as the thick disc (middle panel), the largest fraction originates  in the inner disc (2-4 kpc) where most stars are formed quite early; at such early times, there are very few stars formed  outside 4-5 kpc, as can be seen from the adopted infall timescales in the top panel of Fig. \ref{fig:f1_discs_gen}. Of particular importance is the resulting distribution for the stars in the  Solar neighborhood (red histogram in the middle panel of Fig. \ref{fig:f2_discs_gen}):  an increasingly large fraction comes from the inner disc and $\sim$ 15-20\% of all local old stars are formed at R$_G\sim$3-4 kpc. This is in excellent agreement, both qualitatively and quantitatively, with the findings of \citet[their Fig.3, left panel]{Minchev2013} or \citet[their Fig.10, 3d column]{Sharma2021}, where they display the fractional contribution of various birth radii to the Solar neighborhood population of stars older than 10 Gy. In contrast, the outer disc receives only a small contribution of its old stars from the inner disc (green histogram in the middle panel). 

On the other hand,  most of the "young" stars ($<9$ Gy, thin disc, bottom panel in Fig. \ref{fig:f2_discs_gen}) are formed close to their present position, but they may receive substantial contributions from regions further away, mostly from the inner disc. This is particularly true for the Solar neighborhood in a Galactocentric distance of R $=8$ kpc (red histogram), which receives a contribution from stars born at R$_{\rm{birth}} =3$ kpc. As noticed long ago \citep[e.g.][etc]{chi09,Minchev2013}, this contribution from the inner disc explains the existence of super-Solar metallicity stars in the Solar neighborhood today (see next subsections).

The results regarding the radial displacement of the disc stars of our model are summarized in Fig. \ref{fig:f2_Rb_vs_Rf}, displaying the distribution of birth radii vs final radii for all the star particles of the simulation. Outside  R$_G=3$ kpc, the average birth radius of "young" stars (blue curve) is found to be up to $\sim 2$ kpc inwards of their present day position. In particular, the average birth radius of Solar neighborhood stars is at R$_G=6$ kpc, again in agreement with e.g. \cite{Minchev2013} or \cite{Kubryk2015a}. For "old" stars, the average birth radius (red curve) is found further inwards, as discussed in the previous paragraph. 
As found in \cite{Beraldo2021}, stars can
visit the Solar Neighbourhood by oscillating around their guiding
radius (blurring, most relevant for eccentric orbits) or via radial
migration (churning). 

\cite{Feltzing2020} performed a bold and thorough investigation  of the amount of radial migration in the MW disc, using data for red giant branch stars from APOGEE DR14, parallaxes from Gaia, and stellar ages based on the C and N abundances. In order to estimate the birth radii of the stars, they used  results of the ISM abundances in various epochs, as obtained in the models of  \cite{Minchev2018}, \cite{Frankel2018}, \cite{Sanders2015} and \cite{Kubryk2015a}. They evaluated the fractions of stars that moved radially, either through blurring or churning and found that  half of the stars have experienced some sort of radial migration, 10 per cent likely
have suffered only from churning, and a modest 5–7 per cent have never experienced either churning or blurring. They found that their results depend little on the radial abundance profiles of the adopted models, despite the differences among the latter. They also found that the Sun likely formed between 5.5 and 7 kpc from the Galactic centre in all the aforementioned simulations, again in agreement with our results in the previous paragraph; see, however, next subsection for a smaller Solar R$_{\rm B}$ invoked by \citet[]{Lu2022b}.

\subsection{Age vs metallicity in the local disc}
\label{sub:Local_AgeVsZ}

\begin{figure}
	\includegraphics[width=0.49\textwidth]{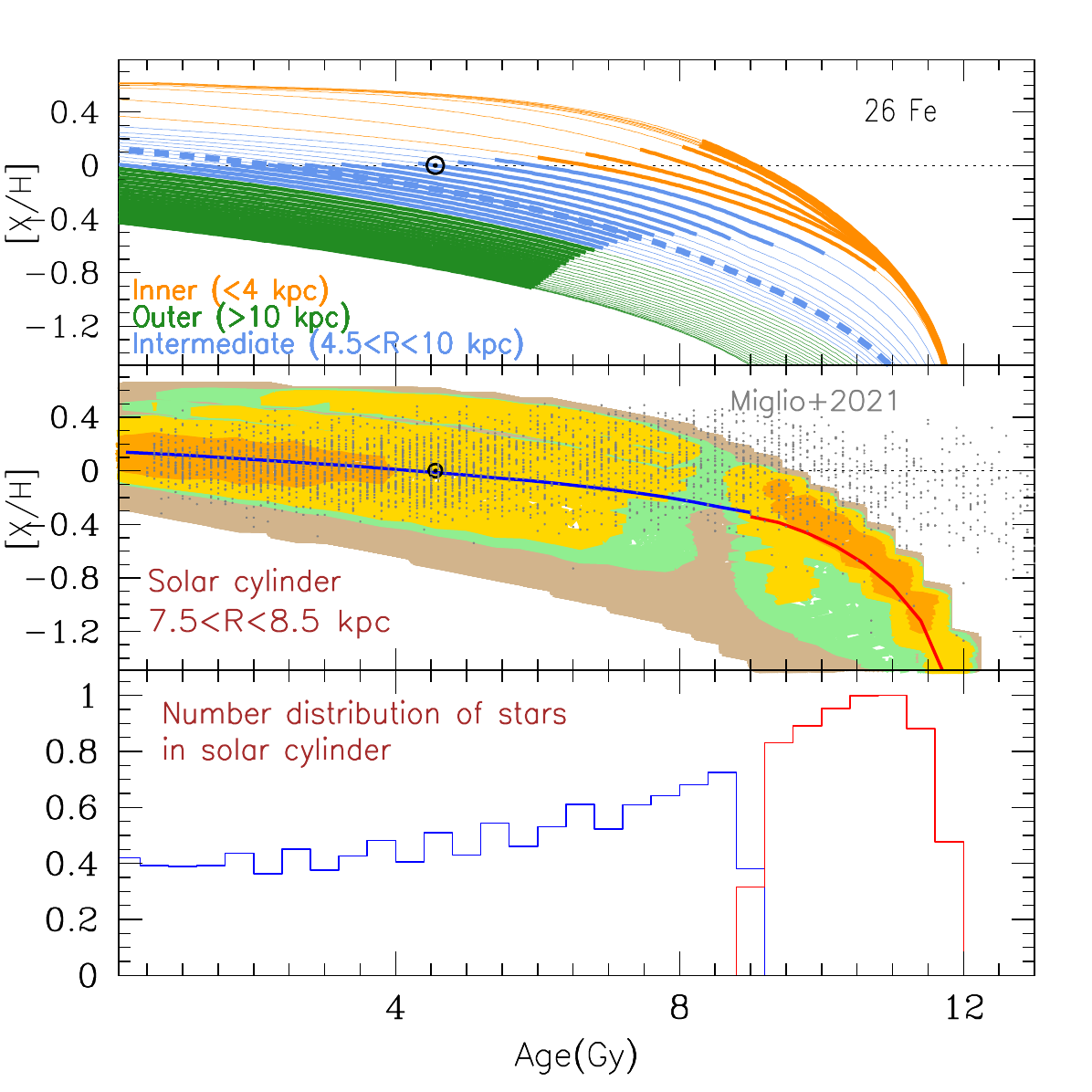}
    \caption{ [Fe/H] vs age. {\it Top:} Evolution of [Fe/H] in the gas in each radial zone. 
    Inner disc ones ($R<4$ kpc) are in red, outer disc ones ($R>10$ kpc) in green and intermediate ones in blue; 
    the dotted curve corresponds to $R_\odot=8$ kpc. The {\it thick} portions of each curve correspond to a local 
    star formation rate higher than the average one for that zone (see text). 
    {\it Middle:} [Fe/H] of stars present in the Solar cylinder ($7.5<R$ (kpc) $<8.5$) at the end of the simulation. 
    Isocontours correspond to star number counts (colour coded as in Fig. \ref{fig:f2_Rb_vs_Rf}), while the blue and red curves indicate the average [Fe/H] for young ($<$9 Gy) and old ($>$9 Gy) stars,
    respectively, and data from \citep[from][]{Miglio2021} are displayed as grey points (see text). 
     {\it Bottom:}      Number distribution of stars present today in the Solar cylinder. 
  }
    \label{fig:f3_FeH_vs_Age}
\end{figure}

Fig. \ref{fig:f3_FeH_vs_Age}  displays the evolution of [Fe/H]. The upper panel shows the evolution in the gas of all the radial zones of the simulation. The  zones  are colour-coded according to their position: red (inner disc, with same short  infall timescales of $\sim$1 Gy), green (outer disc, same large infall timescales of $\sim 8$ Gy) and blue (intermediate disc, with widely  varying infall timescales). The corresponding star formation activity in the various zones is depicted by the thick portions of the curves, denoting star formation rates higher than the average of each zone. For the inner zones, that period occurs early on (in the first $\sim 4$ Gy), while for the outer zones star formation activity is strongest in the last $5-6$ Gy and an intermediate situation holds for the intermediate (blue) zones. In all the cases, the star formation activity is never high at superSolar metallicities, which suggests a rather small fraction of Galactic stars with such metallicities.

%\begin{figure}
%	\includegraphics[width=0.49\textwidth]{f_grad_time.pdf}%{a1_new.pdf}
%    \caption{Evolution of [Fe/H] gradient vs age, as observed in Open Clusters  of ages in the 0-6 Gy range  by Gaia DR3 \citep{Recio-Blanco2022} ; they are compared to model results for stars in the local disc (brown solid curve), taking into account radial migration. Gradients for both data and model curve are for current Galactocentric radius and the latter is evaluated for age bins of 1 Gy. }
%    \label{fig:FeH_gradient_Age}
%\end{figure}

Solar metallicity is reached only fairly recently in the gas of the Solar neighborhood in our model. This is in agreement with observations of local gas, showing that the abundances of several elements within $\sim 1$ kpc from the Sun are Solar to within $\pm 0.04$ dex \citep{Cartledge2006}. It is also in agreement with observations of young B-stars both in the field and in the nearby star forming region of Orion \citep{Nieva2012}, displaying Solar abundances. The straightforward implication of those observations is that, either a) the Sun was formed in its current Galactocentric radius of R$_G=8$ kpc and the local gas metallicity has varied very little since then, or b) the Sun was formed in the inner disc, a couple of kpc inwards from its inner position. The former option implies that, for about half the age of the thin disc, there is an implausibly  perfect equilibrium between metal enrichment of the gas from various nucleosynthesis sources (even those evolving on different timescales, like $\alpha$-elements from massive stars and Fe from SNIa) and metal dilution (from ejecta of metal poor old low-mass stars or late infall). Option (b) implies that the Sun migrated to its present-day position from its birth place, located a couple of kpc inwards in the disc \cite[e.g.][]{Nieva2012,Minchev2013,Kubryk2015a}. This is what we found also here: [Fe/H] becomes zero 4.56 Gy ago in the gas of the zone at Galactocentric radius  R$_G\sim6$ kpc.

The middle panel of Fig. \ref{fig:f3_FeH_vs_Age} shows the situation for the stars now present in the Solar neighborhood, taken as an annulus of width $\rm \Delta$R$=1$ kpc centered at Galactocentric radius R$_0=8$ kpc. We also display the data of \cite{Miglio2021}, showing a disagreement with the model for the oldest ages, which is explainable by the still important uncertainties  (of $\pm$20\% at least) currently affecting the evaluation of stellar ages. The coloured iso-density contours of the model  show two regions of enhanced stellar populations: an early one in the first $\sim 3$ Gy with rapidly increasing [Fe/H] and a late one,  extended over the past $\sim 7$ Gy with very slowly increasing [Fe/H]. Comparison to the upper panel suggests that the early enhancement results from stars formed mainly in the inner disc in the first $\sim 3$ Gy and brought in the Solar neighborhood due to the assumed high radial velocity dispersion at early epochs. The late enhancement of stellar population with [Fe/H]$\sim$0 dex corresponds to stars formed at late times mostly locally, i.e. in the 7-9 kpc region\footnote{We note that both the middle and bottom panels concern the stars that are alive today and not all the stars ever formed. This means that their numbers depend on the SFR of the various radial zones, but for older ages, the upper part of the IMF is missing; e.g. for a 11 Gy old population, only stars below 0.9 \ms \ currently remain, i.e. $\sim$30 \% of the initial mass of stars formed in that period and 10 \% of the initial number are missing.}.

The gap between the two densely populated regions, namely  the paucity of stars around 8-9 Gy is due to the combination of two factors: a) our assumption of  a very rapid and efficient early star formation in the inner disc (due to the very short infall timescale of $\sim$1 Gy for R$_G<4$ kpc), while at larger radii stars are formed at a slower rate ; and b)  the assumption of a highly turbulent gaseous disc in the first 3 Gy, which makes possible the presence of a large number of old stars from the inner disc at 8 kpc.
Stars from intermediate regions (4 to 8 kpc) are also present in the Solar neighborhood now, but they have been formed with largely different star formation efficiencies and their \feh  \ distribution (characterising the metallicity dispersion at a given age) is fairly wide at ages 8-9 Gy; it becomes less wide at younger ages, because younger stars have less time to migrate up to the Solar vicinity. In summary,  the Solar neighborhood is found to be less densely populated with stars  8-9 Gy old, an age which marks the change in the regime of both radial displacement and time-scale of star formation.

We note that similar well-defined overdensities in the  age-metallicity plane are found in the recent analysis of $\sim$250 000 subgiant stars from LAMOST DR7 spectroscopic survey and Gaia EDR3  mission by \cite{XiangRix2022}. The two overdensities are separated by a gap at age 8-9 Gy, as in our case. However, the ages of the oldest stars in \cite{XiangRix2022} extend up to 15 Gy (instead to 12 Gy in our case), making the duration of the thick disc phase much longer than usually assumed. 

Those two enhancements of the stellar density in the age-metallicity plane, are reproduced also in the phase-space of other observables, like the abundance metallicity ratios, as we shall see in subsequent sections. This concerns in particular the \afe \ vs [Fe/H] \ relation and the chemical differentiation of the thin and thick discs.

In the bottom panel of the same figure, we display the number distribution of stars present today in  the Solar neighborhood. Since our model is a 1D and assumes azimuthal symmetry (no vertical dimension), it is more appropriate 
to refer to this region as Solar cylinder. We emphasize that this distribution is not representative of the local star formation rate, since the fraction of migrated stars increases with age; this concerns  stars born elsewhere that migrated to the Solar cylinder, as well as  stars formed locally that have migrated to other zones.

\begin{figure}
	\includegraphics[width=0.49\textwidth]{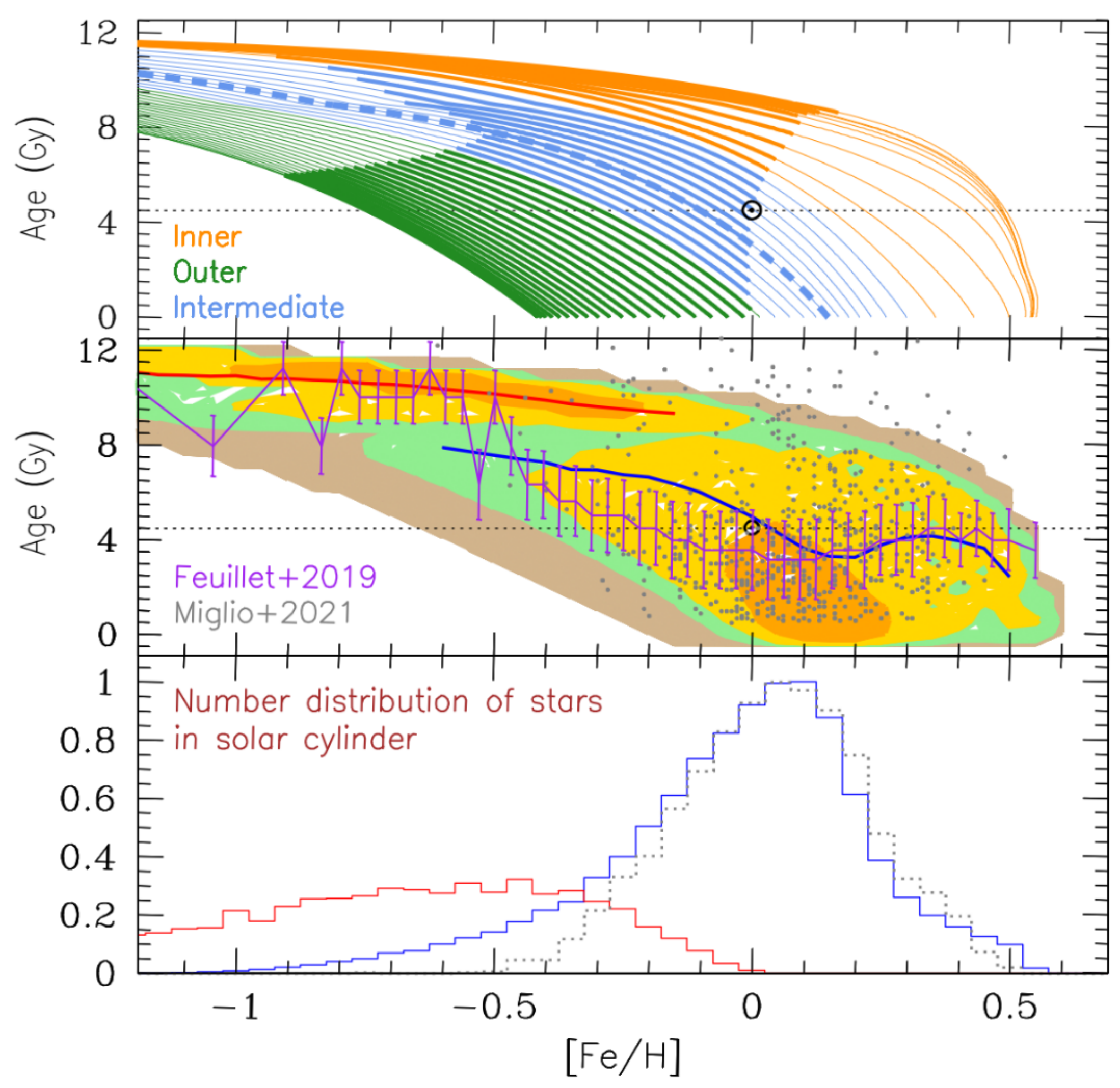}
	%{f5_latest.pdf}%{a20_new.pdf}
    \caption{Age vs metallicity. {\it Top:} Relation for the radial zones of the model, colour code as in Fig. \ref{fig:f3_FeH_vs_Age}. {\it Middle:} Isodensity contours in the Age vs \feh \ plane in the Solar cylinder, colour coded as in
Fig. 3, while the blue and red curves indicate the average [Fe/H] for "young"
and "old" stars, respectively; they are compared to observational data from \citet[violet curve,vertical bars]{Feuillet2019}  and \citet[grey points]{Miglio2021}. {\it Bottom:} Metallicity distributions for Solar neighborhood for thick (red) and thin (blue) discs; the latter is compared to red giant data from \citet[grey dotted histogram]{Miglio2021}. }
\label{fig:f30_Age_vs_FeH}
\end{figure}

Fig. \ref{fig:f30_Age_vs_FeH}
displays another important feature of disc models with radial migration. The previous relation is now shown as age vs metallicity. In the middle panel, the model results show again the double-branch behaviour already discussed in the previous paragraphs. The results are compared to those of \cite{Feuillet2019}, who used a sample of 81,000 stars from SDSS and APOGEE data with Gaia DR2 parallax measurements to derive the age metallicity relation  in several regions of the Galactic disc; their study covered the inner, local  and outer disc, as well as the regions near the plane and away from it. The results shown are those for $7<\rm R(kpc) <9$ and  $0<\rm z(kpc) <0.5$ in Fig. 3 of  \cite{Feuillet2019} (shown with vertical bars and corresponding average values in our Fig.  \ref{fig:f30_Age_vs_FeH}).

At low metallicities ([Fe/H]$<-0.5$) the data suggest a plateau for old ages, which corresponds well to the flat shape of the upper overdensity of our model; these are obviously stars of the "old" (=thick) disc. At higher metallicities, the average age of the observations declines slowly with increasing metallicity, reaching a value of $\sim 3$ Gy at [Fe/H]$\sim 0$, above which it increases again and reaches a plateau around 4 Gy at [Fe/H]$\sim0.5$. As noticed in \cite{Feuillet2019}, the age upturn  at $\sim$Solar metallicity was predicted by the model of \cite{Kubryk2015a} and it is a clear sign of radial migration. While young stars with [Fe/H]$\sim 0$ are formed in the Solar ring (and adjacent regions), higher metallicity stars are formed in the inner disc and transported in our vicinity by radial migration
(see upper panel). As discussed in  \cite{Kubryk2013}  this process proceeds at an average speed of $\upsilon_r\sim1$ kpc/Gy in the radial direction. Stars with superSolar metallicities  have been formed at Galactocentric radii of 3-6 kpc and have travelled to our vicinity on time scale of $\sim \Delta R/\upsilon_r\sim 4-5$ Gy. Our results in this work reproduce quite well the observations (the decline and rise of the average age as function of metallicity) although the model average age is by $\sim 1$ Gy older. A further sign of downturn is observed in the data as well as in our model around [Fe/H]=0.3-0.4 dex.   \cite{Feuillet2019} argue that this feature represents the few most metal-rich stars that were formed rather recently in the innermost disc (R$<2-3$ kpc) and had time to migrate to the Solar neighborhood. Our results (upper and middle panel in Fig. \ref{fig:f30_Age_vs_FeH}) confirm their conclusion. 

Similar results are obtained in the recent study of \cite{Johnson2021} who use a "hybrid" model: radial migration with positions and kinematics of star particles is  taken from a N-body+SPH code, as in \cite{Minchev2013}, and different  star formation histories are explored. As in \cite{Kubryk2015a}, the stars may release their nucleosynthesis products away from their birthplace \footnote{This is an important implication of radial migration for long-lived nucleosynthesis sources, as first pointed out in \citet{Kubryk2013}. }. \cite{Johnson2021} obtain a similar age vs metallicity relation (see their Fig. 16 bottom) to the one displayed
in our Fig.  \ref{fig:f30_Age_vs_FeH} (middle): slightly larger average ages than the observations of \cite{Feuillet2019} at sub-Solar and super-Solar metallicities and approximately the same average age as in \cite{Feuillet2019} at Solar metallicity;  most importantly, the U-turn shape around \zs \ (indicative of radial migration as discussed in the previous paragraph) accompanied by a downturn at the highest metallicities. Despite the very different assumptions of the two models, the similarity of the results between the two studies and the observations suggest that the adopted inside-out scheme of star formation, accompanied by radial migration,  provides a rather realistic description of the evolution of the thin disc.

In a recent work, \cite{Dantas2022}  report the identification of a set of old super metal-rich (+0.15 $\leq$ \feh \ $\leq$ +0.50) dwarf stars with  low eccentricity orbits
 that reach a maximum height from the Galactic plane between $\sim$ 0.5-1.5 kpc and have median ages 7-9 Gy. They  suggest that most  stars in this population originated in the inner regions of the Milky Way (inner disc and/or bulge) and later migrated to the Solar neighbourhood, mostly through churning. An inspection of our  Fig. \ref{fig:f30_Age_vs_FeH} shows that, in our model, such stars would be formed in  Galactocentric distances around 2-3 kpc during the transition phase from thick to thin disc. They are considerably older (by several Gy) than most stars of similar metallicities currently present in the Solar neighborhood. The detailed analysis of their abundance patterns, which are presented in \cite{Dantas2022}, would provide further  insight  into the nucleosynthesis in those remote Galactic regions and will be the topic of future work.

\begin{figure}
	\includegraphics[width=0.49\textwidth]{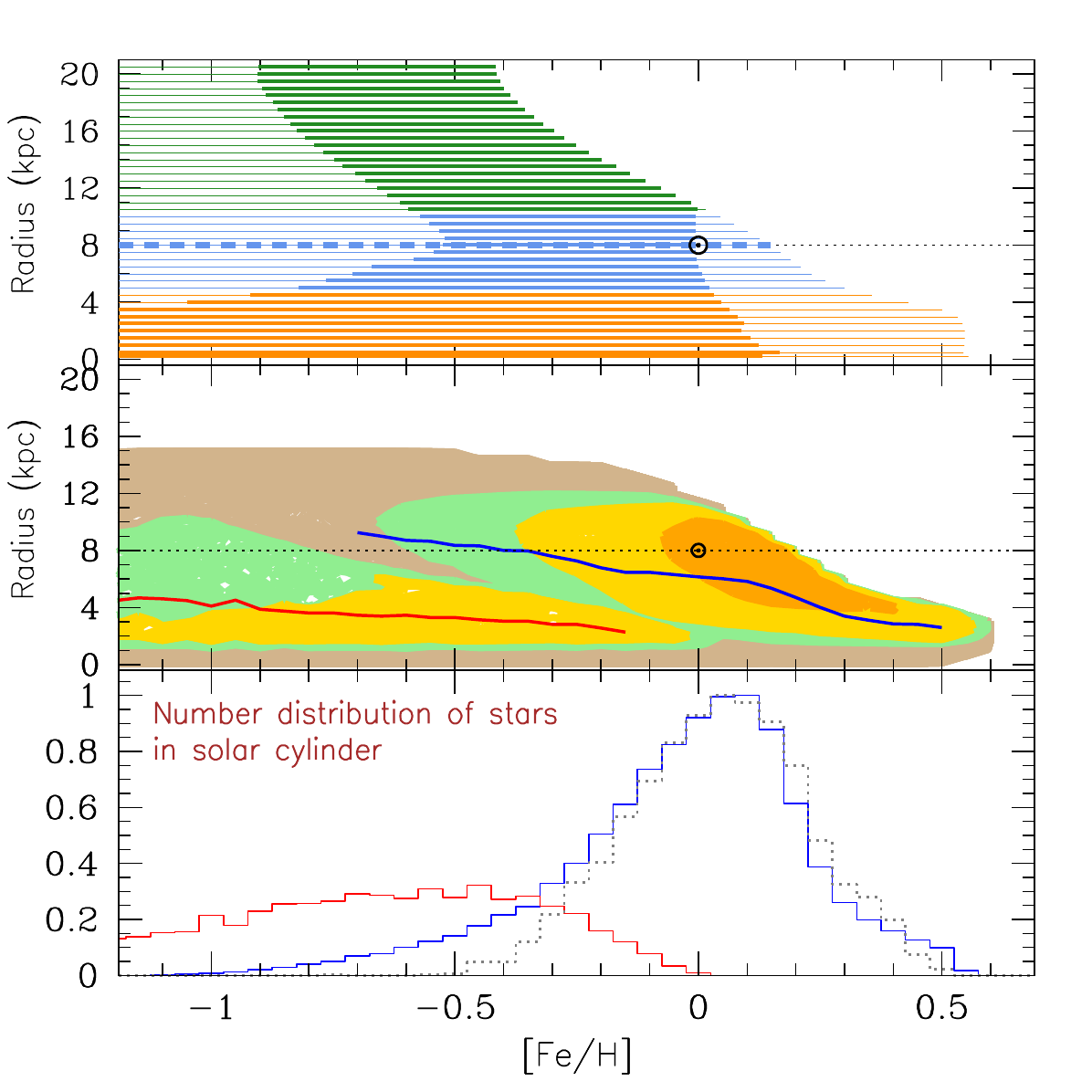}
    \caption{Birth radius vs stellar metallicity. {\it Top:} Evolution of gaseous [Fe/H] in all the radial zones of the model, with colour coding as in Fig.\ref{fig:f3_FeH_vs_Age}. {\it Middle:} Isodensity contours for the birth radii of stars present today in the Solar cylinder, colour coded as in
Fig. 3, while the blue and red curves indicate the average [Fe/H] for "young"
and "old" stars, respectively ; the thick solid curves indicate averages at a given metallicity for the old ($>$9 Gy, red) and young ($<$9 Gy, blue) discs. {\it Bottom:} Metallicity distributions for Solar neighborhood for thick (red) and thin (blue) discs; the latter are compared to red giant data from \citet[grey dotted histogram]{Miglio2021}.  }
   \label{fig:BirthRad_vs_FeH}
\end{figure}

In the lower panel of Fig. \ref{fig:f30_Age_vs_FeH} we display the metallicity distribution of stars for the thin and thick discs in the local cylinder. We compare the one for our thin disc (all stars with ages $<$9 Gy) with   data for the thin disc obtained recently by \cite{Miglio2021} 
for  3,300 giant stars with available Kepler light curves and APOGEE spectra. In that study, stellar masses and ages are obtained from a combination of seismic indices
and photospheric constraints, and the authors emphasize that those properties depend on the observational constraints via power
laws, which lead to  "{\it a blurred view at older stellar ages"}. For that reason, and because that survey obtains quite old ages (up to 14 Gy), which are substantially higher than the 12 Gy of our oldest stars, we chose to compare our results only with the thin disc data of \cite{Miglio2021}, which we consider to correspond to stellar age $<$9Gy(in agreement with our definition). In the bottom panel of Fig. \ref{fig:f30_Age_vs_FeH} we find a fairly good agreement in most of the metallicity range between model and data, except in the lowest metallicities where our model displays a rather extended tail down to \feh$\sim$-1, while the \cite{Miglio2021} thin disc does not go below \feh$\sim$-0.5.

Fig. \ref{fig:BirthRad_vs_FeH} illustrates clearly the spatial origin of the stars present today in the Solar neighborhood for the various metallicity ranges. The star formation activity throughout the disc has been more intense at sub-Solar metallicities (top panel), although this occurred at fairly early times in the inner zones and at late times for the outer zones, as discussed in previous paragraphs. The bulk of the stars in the Solar cylinder is around \feh$\sim0$ and originates from the $5-9$ kpc region. The high metallicity tail of the distribution of stars younger than 9 Gy(blue histogram in lower panel) originates from zones at the interior of 4 kpc, and on average in the zone at R$=3$ kpc (blue curve in middle panel). The low metallicity tail of the young (=thin) disc has an important contribution from the outer zones at $9-10$ kpc, as already suggested in \cite{Haywood2008} and \cite{Schon2009}, on the basis of observational and theoretical analysis, respectively. The old(=thick) disc extends up to quasi-Solar values and its stars originate mostly from radii $<6$ kpc  as already illustrated in Figs. \ref{fig:f2_discs_gen} and \ref{fig:f2_Rb_vs_Rf}.
This is also valid for the superSolar metallicity stars of the thin disc, as already discussed in the previous paragraphs and in \cite{Dantas2022}.

\subsection{Abundance profiles and their evolution}
\label{sub:Abundance_profiles}

Abundance profiles constitute a major probe of the physics of the disc evolution, since they are connected to the radial dependence of several key ingredients, like star formation, infall and stellar yields, and various tracers have been explored 
(\cite{Luck2011,Magrini2017,Magrini2023})

\begin{figure}
	\includegraphics[width=0.49\textwidth]{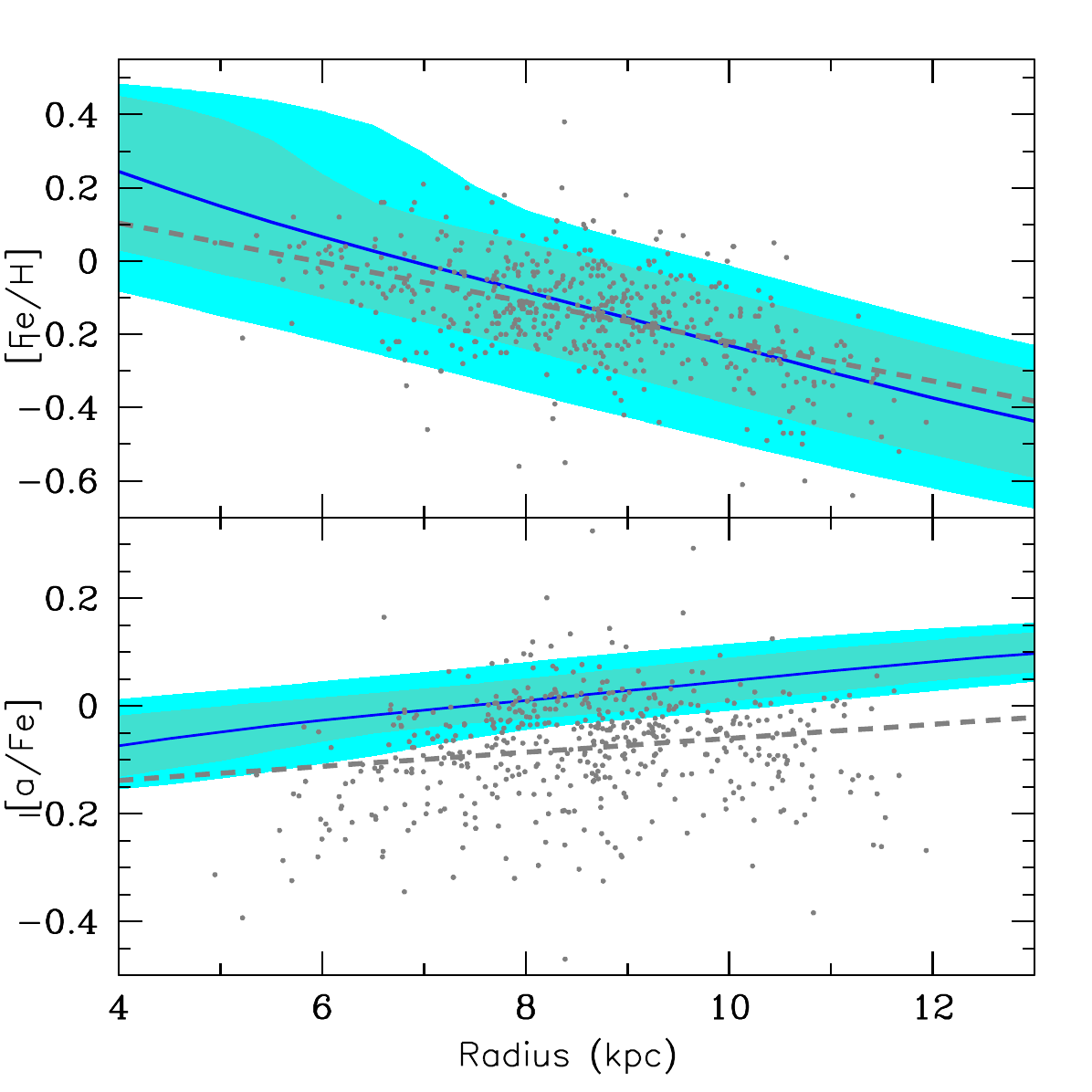}
    \caption{ {\it Top :}   Metallicity profile of open clusters (ages in the 0-6 Gy range) as observed by Gaia DR3 \citep{Recio-Blanco2022b}, with the average value indicated by the grey dashed curve; they are compared to model results (blue curves), with the  shaded areas indicating the 1-sigma and 2-sigma density contours. {\it Bottom:} Same as in previous panel, for the \afe\  profile. 
    }
    \label{fig: ModelVsGAIA_profiles}
\end{figure}

The abundance profiles of our model appear in Fig. \ref{fig: ModelVsGAIA_profiles}, where they are compared to the recent data of Gaia DR3 \citep{Recio-Blanco2022b},  based on Gaia/RVS chemical data from \cite{Recio-Blanco2022}. The data concern open clusters with ages $<$5.5 Gy found today in the Galactocentric radial range R$_G=5-11$ kpc. Our model results are displayed for the same radial and age ranges. We notice, however, that the data do not correspond exactly to the same quantities, although sometimes such "proxies" are used in the literature. Thus, we provide [Fe/H] values, whereas Gaia data concern [M/H] values (M referring to the average abundance of several iron-peak elements), and we use Si as proxy for the $\alpha$ elements\footnote{The evolution of [Si/Fe] vs [Fe/H] is well reproduced in the 1-zone model of the Solar neighborhood of \cite{Prantzos2018} with the yields adopted here.} provided by Gaia DR3.  Furthermore, our mean values are for all stars younger than 5.5 Gy found in a given radial bin, while the Gaia DR3 mean values depend obviously on the ages of the observed open clusters found in each radial bin, which are on average considerably lower than 5.5 Gy.

In the top panel of Fig. 7, the average value of the model \feh \  in  the Solar vicinity (R$=8$ kpc) is slightly  higher than the one of the [M/H] of the observational data. Also, the slope of the  metallicity profile of the model, d\feh/dR$=-0.069$ dex/kpc (solid curve) is slightly steeper than the data slope of d[M/H]/dR$=-0.054\pm0.008$ dex/kpc (dashed curve). Both values are consistent with those recently reported in  other surveys of open clusters \citep[see references in][]{Recio-Blanco2022b,Spina2022}, albeit with much smaller samples. The 1$\sigma$ and 2$\sigma$ ranges of the model [Fe/H] values (shaded aereas) include most of the data points. 

In the bottom panel of Fig. \ref{fig: ModelVsGAIA_profiles}
are displayed the corresponding \afe  \ ratios for the same age and radial ranges. Here the slope of the model profile (+0.014 dex/kpc) is in excellent agreement with the one of the data ($0.013\pm0.007$ dex/kpc), but there is  an important overall offset of 0.1 dex: the model results an average value of \afe$\sim$0 in the Solar vicinity, whether Gaia DR3 data analysis suggests a value 0.1 dex lower. 

At this point, we note that the profiles presented in Fig. \ref{fig: ModelVsGAIA_profiles} correspond to all the stars of our model, whereas Gaia DR3 data concern open clusters. It is not clear whether such massive objects are affected to the same extend as single stars by radial migration, so a direct comparison may not be appropriate. Indeed, open clusters from the inner disc, interacting strongly with inhomogeneities of the gravitational potential, may dissolve in rather short timescales and not undergo the same amount of radial migration as field stars \citep{Anders2017,Spina2021}. Thus, despite the precision with which their ages, Galactocentric distances and chemical  abundances are measured, the comparison with our model is not straightforward.

The evolution of the abundance profile has been a key issue in studies of the MW disc. Most models predict a flattening of the abundance profile of the disc gas with time and thus a steepening of the corresponding profile of stars with age, although few ones predict the opposite effect 
\citep[see][and references therein]{Hou2000,Pilkington2012,Molla2019}. However, the stellar abundance profile of the disc, as observed today, can be altered by radial migration, as already noticed in e.g. \cite{Minchev2013,Minchev2014} or \cite{Kubryk2015b}.

\begin{figure}
	\includegraphics[width=0.49\textwidth]{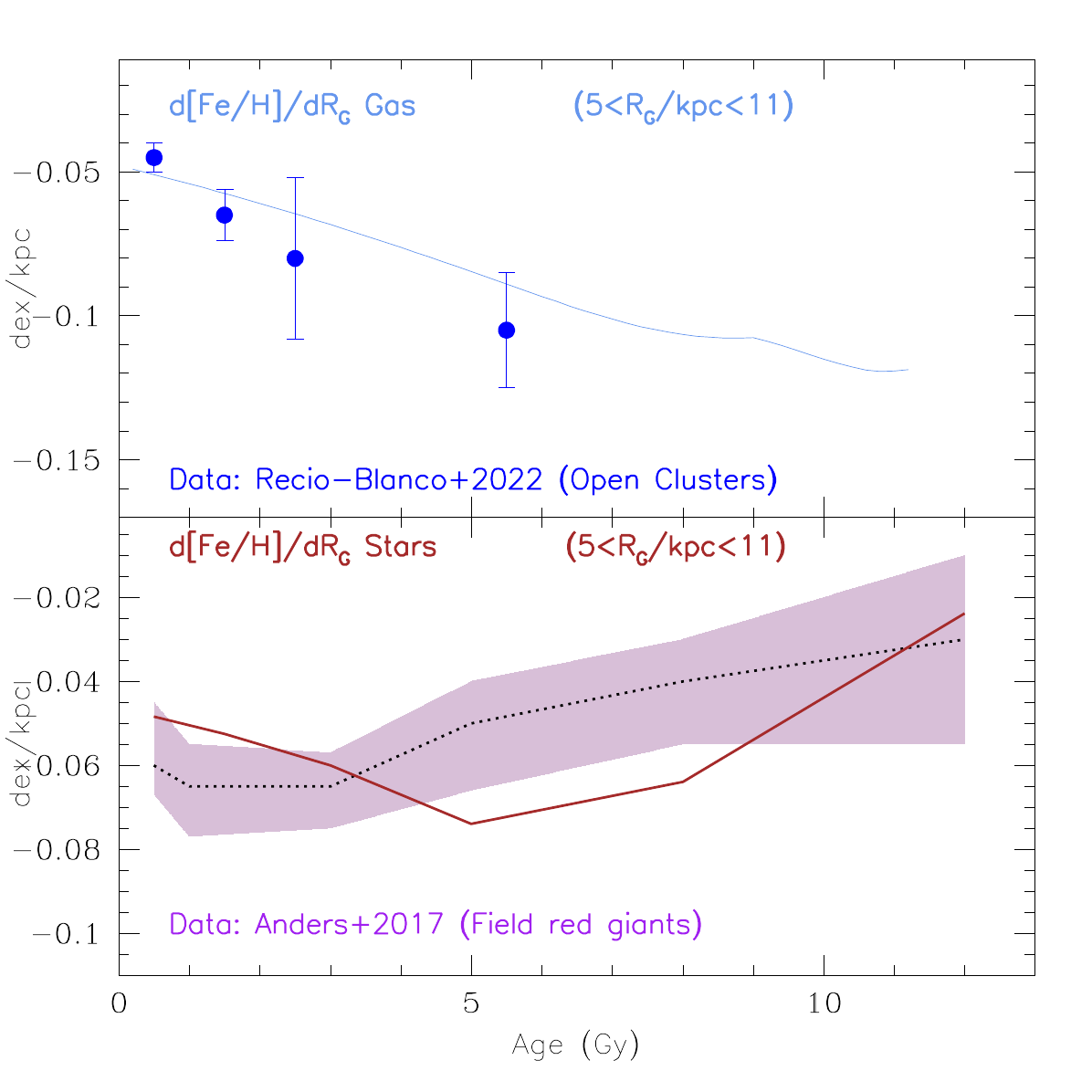}
	%{f_evol_grad0.pdf}
    \caption{   Evolution of abundance gradient $\Delta$\feh/$\Delta {\rm R_G}$  (solid curves) in the range of  Galactocentric radii R${\rm _G}$=5-11 kpc, in the gas (top) and in the stars (accounting for radial migration, bottom), and comparison to observations: Open clusters  \citep[top, blue dots from][]{Recio-Blanco2022b}; Field red giants \citep[bottom, dotted curve with uncertainties within the shaded aerea from][]{Anders2017}. The age bins adopted in the latter study are also used to evaluate the stellar gradient of the model.}
    \label{fig:ModelvsAge_gradients1}
\end{figure}

 In the top panel of Fig. \ref{fig:ModelvsAge_gradients1} we show the evolution of the  \feh \ gradient in the gas, in the range of Galactocentric distance R$_G$ from 5 to 11 kpc which decreases from ~-0.05 dex/kpc today to ~-0.12 dex/kpc $\sim$11 Gy ago, i.e. with a rate of -0.007 dex/kpc/Gy.   Blue symbols with error bars correspond the  Gaia DR3 analysis of \cite{Recio-Blanco2022b} for open clusters younger than 5.5 Gy present in the same fixed range of Galactocentric radii. Despite the large error bars, the gradient clearly decreases with age, at a rate slightly larger than in our model.  As emphasized before, the variation of the open cluster properties during radial migration do not allow for a straightforward comparison with our model results.  
%The corresponding gradient of our model, evaluated in the same conditions, i.e. for all stars younger than 5.5 Gy now present in the same radial range (blue solid curve),  is compatible with the observed one - within error bars - although it steepens at a slower rate. We note, however, that both observed and model gradients in this case are affected by radial migration and may not reflect the evolution of the true abundance gradient, which is the one of the gas at the time of birth of a given stellar population.

In the bottom panel of Fig. \ref{fig:ModelvsAge_gradients1} we show the evolution of the  \feh \ gradient in stars, as obervable today in the same range of Galactocentric distance R$_G$ from 5 to 11 kpc. As already shown by \cite{Minchev2013} or \cite{Kubryk2015b}, the stellar gradient is affected by radial migration and is flattened with respect to the gaseous one, because of the radial mixing of stellar populations. The effect is amplified with stellar age, since older stars have more time to migrate. This is seen in the CoRoT+APOGEE data analyzed by \cite{Anders2017}, showing a slight early decrease of the slope for young stars, followed by a steady increase after a few Gy. 

The corresponding behaviour of our model gradient is qualitatively similar, but the decrease period lasts $\sim$5 Gy, taking the stellar gradient to  values lower than the observed ones in that period. The subsequent rise of the gradient is also seen but it is faster than in the observations in the last age bin. In fact, the evolution of the stellar gradient in Fig. 7 (middle panel) of  \cite{Kubryk2015b} is in better agreement with the data of \cite{Anders2017}; the two models are similar but differ in the prescriptions of radial migration in the early Galaxy and in the stellar yields,  as presented in Sec. \ref{subsub:StarsGas} and \ref{subsub:Chemistry}, respectively. At this point, we notice that in our model the gradient is evaluated by considering all the stars of a given age that are still alive today, while observations consider red giants; since the fraction of red giants in a given stellar population is a function of age and metallicity, obviously the two gradients displayed in the bottom panel of Fig. \ref{fig:ModelvsAge_gradients1} are not directly comparable. More work in that respect is needed in our modelling.

\begin{figure}
	\includegraphics[width=0.49\textwidth]{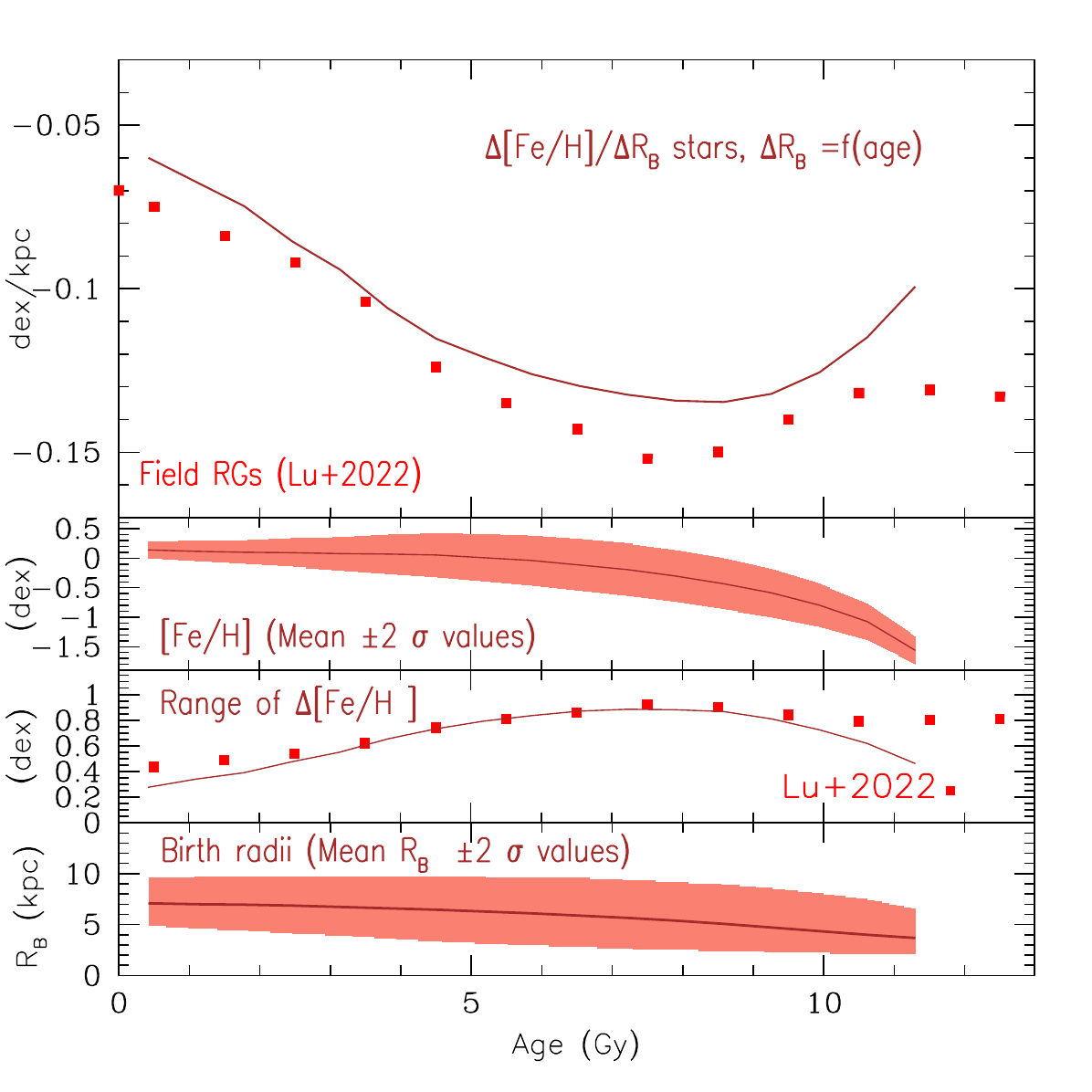}
	%{f_evol_grad0.pdf}
    \caption{ {\it Top:}   Evolution of  abundance gradient $\Delta$\feh/$\Delta {\rm R_B}$ of stars presently found in the Solar neighborhood, evaluated from their birth radii R${\rm _B}$ \citep[red squares from][]{Lu2022b} and corresponding model results (brown solid curve). {\it Second from top:} Metallicities of stars presently found in Solar neighborhood as function of their ages (mean and $\pm$2$\sigma$ values), corresponding to the middle panel of Fig. \ref{fig:f3_FeH_vs_Age}. {\it Third from top:} Range of \feh \ values (in dex)  used to evaluate  the gradient $\Delta$\feh / $\Delta {\rm R_B}$ in the upper panel; data \citep[red points from][]{Lu2022b} and our model (brown curve). {\it Bottom:} Evolution of the range of birth radii $\Delta {\rm R_B}$ of stars presently found in the Solar neighborhood (mean and 2$\sigma$ values); the solid curve corresponds to the average ${\rm R_B}$ value.  The ranges of $\Delta$\feh \ and $\Delta {\rm R_B}$ of the two bottom panels are used to evaluate the $\Delta$\feh / $\Delta {\rm R_B}$ of the upper panel. }
    \label{fig:ModelvsAge_gradients2}
\end{figure}

As already mentioned, radial migration makes it difficult to trace back the evolution of the true abundance gradient in the gas of the disc.
 Based on an idea formulated in \cite{Minchev2018} and explored with numerical simulations in \cite{LuBuck2022}, \cite{Lu2022b} tried recently to circumvent that difficulty.  They use $\sim$80000 subgiant stars of LAMOST presently found in the Galactocentric distance range R$_G$=6 to 12 kpc, with an average age and metallicity uncertaint as small as 0.32 Gy and 0.03 dex, respectively. At each age bin  they take the difference $\Delta$\feh=\feh$_{max}$-\feh$_{min}$ of the most and least metallic stars of their sample {\it born at the given age bin} and they assume that there exists a time-dependent linear relationship between this metallicity range and the metallicity gradient, which they normalize to the present day gradient $\Delta$\feh/$\Delta$R (Age=0)=--0.07 dex/kpc. They find a steepening of the [Fe/H] gradient with age, up to 8-11 Gy ago, at which time the trend is inversed (red squares in top panel of Fig. \ref{fig:ModelvsAge_gradients2}).  \cite{Lu2022b} associate that feature to the last major merger (Gaia Sausage/Enceladus event) and they  claim that this transition plays a major role also in shaping the \afe vs [Fe/H] dichotomy (see Sec. \ref{subsec:afe_discussion}). It is interesting to note that the analysis of \cite{Lu2022b} concludes that the stars formed during that transition period have birth radii around 4-5 kpc, indicating the extend of the disc 8-11 Gy ago;  indeed, the mean birth radius of  the local thick disc stars in our model is around 4 kpc,  as displayed in Fig. \ref{fig:f2_discs_gen} (bottom panel).

Following the work of \cite{Lu2022b}, we evaluate the maximum and minimum metallicity of 95\% of the stars currently present in the Solar neighborhood,  and the corresponding birth radii, as shown in the three lower panels of Fig. \ref{fig:ModelvsAge_gradients2}. The 2nd from top panel displays the $\pm$2$\sigma$ values of metallicity (upper and lower limits of shaded aerea)  around the mean (solid curve). The 3d panel from top panel displays the difference $\Delta$\feh =\feh$_{max}$-\feh$_{min}$ for each age, and we find an excellent agreement with the data points of \cite{Lu2022b}. The bottom panel shows the corresponding range of birth radii 
$\Delta$R$_B$, between the 5th and 95th percentile (shaded aerea). The abundance gradient is then obtained for each age as $\Delta$\feh/$\Delta$R$_B$. We plot  our result (brown solid curve) in the top panel of Fig. \ref{fig:ModelvsAge_gradients2}. The gradient steepens with age in agreement with the results of \cite{Lu2022b}, but it differs  by 0.01 to 0.02 dex/kpc. We note that \cite{Lu2022b} infer the range of birth radii of their LAMOST sample by calibrating it as to get a present day gradient of -0.07 dex/kpc, which its lower than ours by $\sim$0.01 dex/kpc. Had they adopted a different calibration, their data points in Fig. \ref{fig:ModelvsAge_gradients2} (top panel) would be found 0.01 dex/kpc higher and would then be in much better agreement with our model. Moreover, taking into account the various uncertainties involved in the observational estimate of birth radii, discussed in detail in \cite{LuBuck2022}, we find the agreement to be quite satisfactory.

The most important feature, however, is that the model curve also goes through a broad minimum around 9 Gy ago and  then increases by 0.03 dex/kpc at the oldest ages.
This change in the slope of the gradient evolution is obtained in our model because in the oldest ages the birth radii of the stars now present in the Solar neighborhood are mostly located in the inner Galaxy: in those inner regions the abundance profile becomes flat quite early on, since they evolve on similarly short timescales (see 3d panel from top panel in Fig. \ref{fig:f1_discs_gen}). Outside the 4 kpc region, the gradient is quite steep early on and flattens with time. The  stars that we observe locally today is a mixture of different ages and regions:  old stars mostly from the inner disc (small gradient), stars of intermediate ages mostly from intermediate regions (steep gradient) and finally young stars of regions closer to the Solar  neighborhood (again smaller gradient because the abundance profile flattens with time). This evolution of the range of birth radii is displayed in the bottom panel of Fig. \ref{fig:ModelvsAge_gradients2}, where the average birth radius is also shown: it goes from $\sim$4 kpc 11 Gy ago, to $\sim$7 kpc at Sun's birth and $\sim$8 kpc today.

In summary, we find that the sample of stars locally observed is weighted in a time-dependent way that reflects the inside-out formation of the disc (more stars from the inner disc early on). In that respect, we fully agree with the conclusions of \cite{Molla2019}, namely that "{\it the use of a variable radial range, which takes into account the
growth of the disc along the time or redshift, is an important caveat
for estimating the correct evolution of the disc radial gradient.} The  evolution of the stellar abundance gradient observed today results from the synergy of the aforementioned  factors and goes through a minimum which corresponds to the transition, of the thick to the thin disc. However, in our case, this transition does not require some external event, as advocated in \cite{Lu2022b}, but it is made "softly" through secular evolution.

Overall, we may distinguish three types of abundance gradients: \\
a) the "true" abundance gradient in the gas, observable only today \\
b) the abundance gradient in stars, as observed in their current  Galactocentric position; its evolution is observable with sufficiently good stellar ages  \\
c) the abundance gradient that the stars currently observed in some Galactocentric position had at their birth place.

In the absence of radial migration, gradients (a) and (b) evolve in the same way, provided that short age bins are used for (b). Furthermore, {\it if} the gradients are characterized by a unique slope across the whole galactic disc, then gradient (c) also evolves  as gradient (a). With radial migration, the evolution of gradients (a) and (b) is different - gradient (b) being more and more flatter for older ages -, but gradient (c) evolves still the same as gradient (a) {\it in the case of a unique slope over the whole disc} at each time;  this latter property was assumed and explored in the work of \cite{Lu2022b} who inferred the evolution of (a) by reconstructing first the evolution of (c). However, if the slope of gradient (a) is not the same all over the whole disc at a given time, then radial migration makes the evolution of gradient (c) different from both (a) and (b); this last case was explored in this section and illustrated in Figs. \ref{fig:ModelvsAge_gradients1} and \ref{fig:ModelvsAge_gradients2}.

The properties of the Galactic disc discussed in this section, i.e. age and space distribution 
(resulting from the adopted scheme of inside-out disc formation and stellar radial migration through blurring and churning) affect straightforwardly the elemental abundance ratios of the corresponding stellar populations in the Solar neighborhood, as we discuss in more detail  in the next sections.

 %\begin{figure}
 %	\includegraphics[width=0.49\textwidth]{f10_new.pdf}
 %    \caption{Same as in Fig. \ref{fig:f30_Age_vs_FeH}, for age vs [$\alpha$/Fe].   Comparison is made to LAMOST data for red giants \citep{Huang2020} in the middle (grey points) and bottom (grey histogram) panels (see text).}
 %    \label{fig:f10_Age_vs_alphaFe}
 %\end{figure}

\section{The [$\alpha$/Fe] ratios in the thin and thick discs}
\label{sec:alpha_Fe}

\subsection{[$\alpha$/Fe] vs age}
\label{subsec:alpha_FeVsAge}

\begin{figure}
	\includegraphics[width=0.49\textwidth]{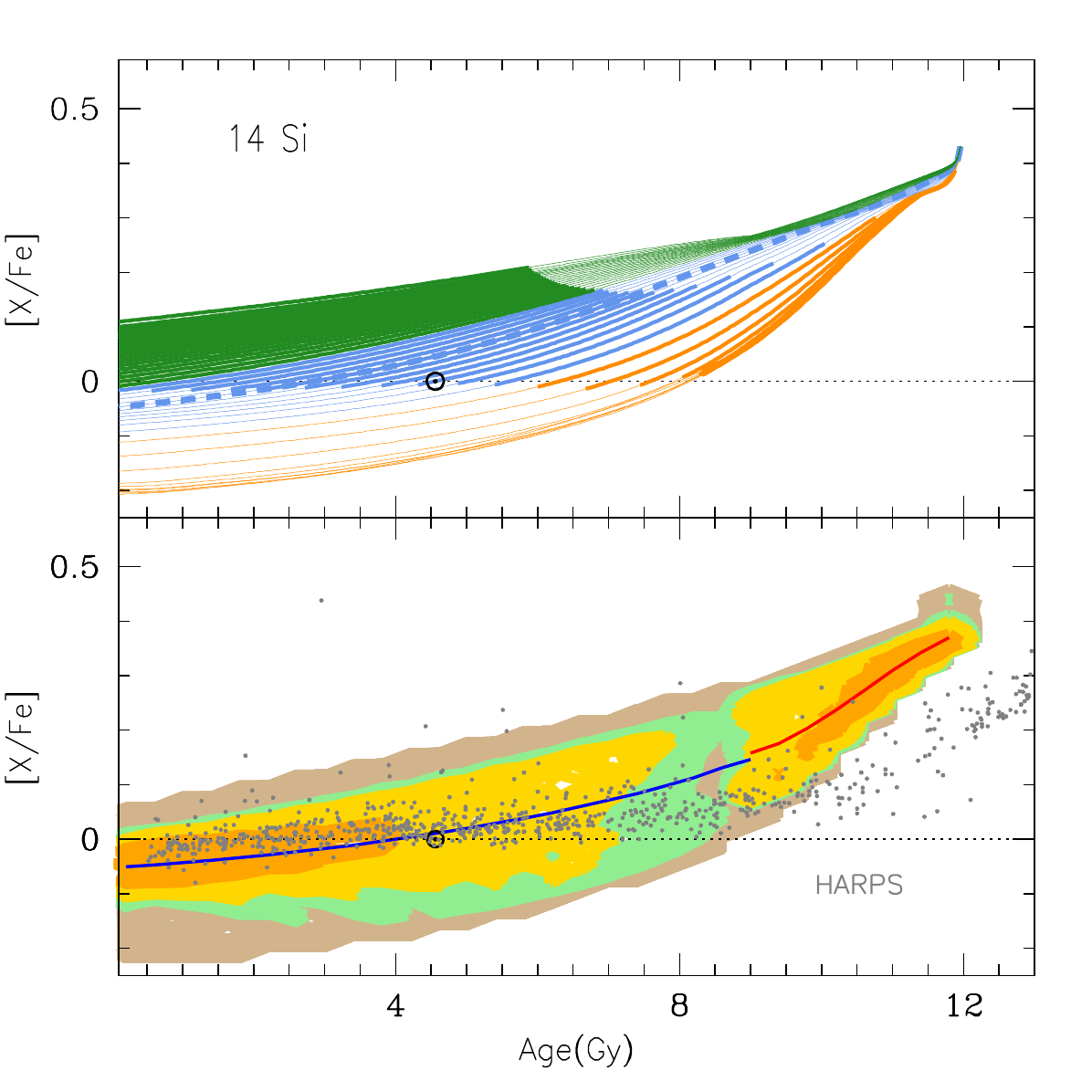}
    \caption{[Si/Fe] vs stellar age. {\it Top:} Evolution of gaseous abundances in all the radial zones of the model, with colour coding as in Fig.\ref{fig:f3_FeH_vs_Age}. {\it Bottom:} Abundances of stars present today in the thin and thick discs in the Solar cylinder (see discussion in the text); abundance data for [$\alpha$/Fe] from HARPS spectra  \citep[grey points,][]{Adibekyan2012,DelgadoM2019}.}
   \label{fig:f4_SiFe_vs_Age}
\end{figure}

Fig. \ref{fig:f4_SiFe_vs_Age} displays the evolution of the ratio [X/Fe] for a typical $\alpha$-element, like Si.
The evolution of [$\alpha$/Fe] depends mostly on the history of star formation in each radial zone (upper panel). [$\alpha$/Fe] decreases monotonically, because of the delayed release of Fe (about 2/3 of Solar Fe) by SNIa. The decrease is more rapid in the inner regions, where the early intense star formation produces a rapid  increase of both CCSN and SNIa, and of the abundances of both Si and Fe, which reach their quasi-equilibrium values within a few Gy. In contrast, the evolution of the outer regions proceeds at much slower pace and a large fraction of their SNIa appears only at late times; as a result, their [Si/Fe] decreases slowly but steadily. The absolute value of the final [Si/Fe] ratio depends on the time integrated rates of CCSN and SNIa. The integral 
$\int_0^{12 Gy}$ R(CCSN)/R(SNIa) $dt$ is smaller in the inner regions than in the outer ones, and so is the corresponding final [Si/Fe]  ratio: it is sub-Solar in the inner zones, but super-Solar in the outer ones.

The evolution of the gas in the Solar cylinder is intermediate between the two extreme regimes, but the stars present today locally are affected in different ways: the stars born at late times (younger than the Sun) have [Si/Fe] ratios similar to those of the gas in the $7-9$ kpc range, while the old stars have ratios similar to those of the gas of the early disc in the $3-4$ kpc range. Thus, the change of the slope of [Si/Fe] vs age of local stars around $\sim$9 Gy marks indeed a transition between two regimes of star formation, but not necessarily because of a "hiatus" in star formation, as sometimes claimed \citep{Snaith2014}: it simply reflects the fact that stars formed in different regions with different (but smooth) star formation histories and different kinematics are found today in the same location. In other terms, the adopted gradual inside-out formation of the disc
provides naturally the transition between the large early slope (dominated by stars from the inner disc) and the almost flat late slope of the thin disc (dominated by local stars). This is also reflected in other chemical properties of the stars in the Solar cylinder, as we discuss in subsequent sections.  But, in any case, a simple inspection of Fig. \ref{fig:f4_SiFe_vs_Age} suggests that, by itself, \afe \ cannot be considered as a reliable proxy for the age of star, although its variation is of course smaller than the one of \feh, which is usually considered as such a proxy (compare the upper panels of Figs. \ref{fig:f3_FeH_vs_Age} and \ref{fig:f4_SiFe_vs_Age}). Finally, as already discussed in the previous sections, the lower panel of the latter figure also shows how a "hiatus" in the number density of stars around 8-9 Gy can be obtained without any pause in SFR or infall in any of the radial zones of the model, 
but simply due to the adopted inside-out scheme of disc formation.

The corresponding data in the bottom panel of Fig. \ref{fig:f4_SiFe_vs_Age} (grey points) from HARPS (High Accuracy Radial velocity Planet Searcher) spectra show indeed a smooth decline of [Si/Fe], but the stellar ages suffer from considerable uncertainties, preventing a meaningful comparison to the old disc of our model.

\subsection{[$\alpha$/Fe] vs metallicity}
\label{subsec:alphaFe_FeH}

\begin{figure}
	\includegraphics[width=0.49\textwidth]{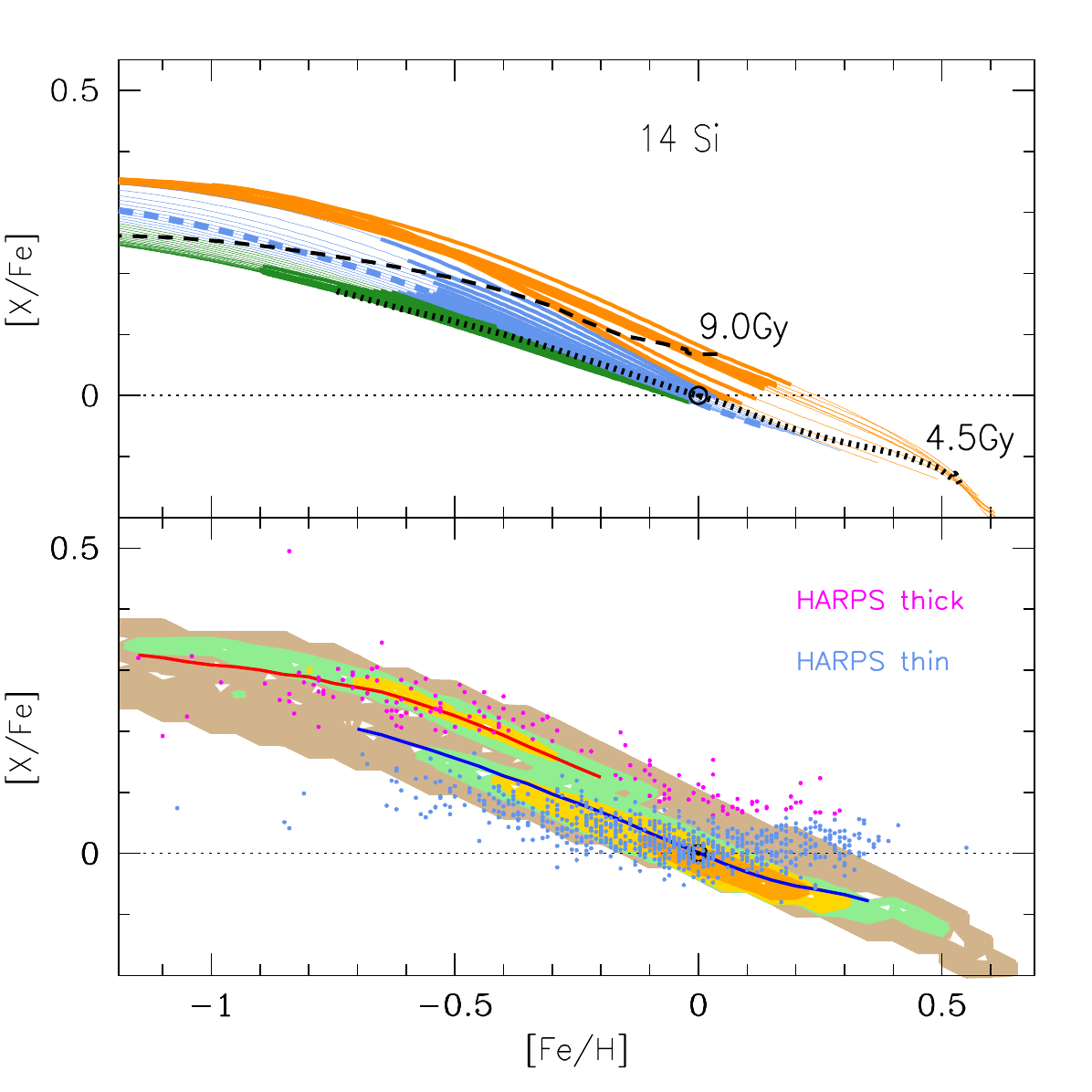} %{a3_new.pdf}
    \caption{Same as in Fig. \ref{fig:f3_FeH_vs_Age}, for [Si/Fe] vs [Fe/H].  Two isochrones are displayed in the top panel, for age equal 9 Gy (dashed curve) separating the "old" from the "young" disc, and for age equal 4.5 Gy  (dotted curve, see discussion in the text). In the bottom panel, two overdensities  appear clearly through the isocontours, corresponding to those appearing in the two previous figures and representing the thick (old) and thin (young) discs. The magenta and blue points are HARPS data for [$\alpha$/Fe] vs [M/H], chemically separated into thick and thin discs, respectively \citep{Adibekyan2012}.}
    \label{fig:f5_SiFe_vsFeH}
\end{figure}
Fig. \ref{fig:f5_SiFe_vsFeH} displays the relation between [Si/Fe] vs [Fe/H], for all the zones of the model (top panel) and for the Solar cylinder (bottom panel). The isochrones in the top panel indicate snapshots at times  of 3 Gy after the beginning of star formation (dashed curve, current age$=9$ Gy) and 8 Gy after star formation (dotted curve, current age$=4$ Gy). The former corresponds to the end of the thick disc formation, while the latter approximately to the period of Sun's formation; at that time, the metallicity in the inner zones was already $\sim2-3$ times Solar and [Si/Fe]$\sim-0.1$,  where in the Solar vicinity (R$_G\sim7-9$ kpc) both quantities were close to Solar.

In the bottom panel, the two sequences corresponding to the thin and thick discs are clearly separated, for the same reason as described for Figs.  \ref{fig:f3_FeH_vs_Age}, \ref{fig:f30_Age_vs_FeH} and \ref{fig:f4_SiFe_vs_Age}: the "gap" in that region of the [Si/Fe] vs [Fe/H] diagram corresponds to a low - albeit not necessarily null - density of star formation (see top panel) and this is reflected also in the stars that populate today the Solar vicinity, after migration.

The results of the model are compared to data from HARPS spectra \citep{Adibekyan2012,DelgadoM2017} for the thin and thick discs. The agreement is quite satisfactory, at least for metallicities up to approximately Solar; for superSolar metallicities, the model [Si/Fe] ratio continues to decrease 
with metallicity, in contrast to observations. The reason is that in the inner disc regions, late SFR is negligible and so is the production rate of alpha elements from CCSN; but SNIa, with progenitors formed in the early period of intense star formation, continue to enrich the ISM with Fe. As a result, the [$\alpha$/Fe] ratio continues decreasing in the inner disc, and some of the stars formed there  are found in the Solar vicinity through radial migration. It is difficult at this point to decide whether the mismatch between model and observations at the highest [Fe/H] values is due to an excessive late SNIa rate in the inner disc, or to an excessive migration of stars from that region into the Solar vicinity, caused by the adopted radial migration scheme. We note that, as shown by \cite{Santos-Peral2020} the flattening of  the \afe \ abundance trend at superSolar metallicities could be an observational artifact due to uncertainties in the continuum placement. A specific normalisation treatment as that of \cite{Santos-Peral2020}can help to recover a decreasing trend with metallicity  in agreement with our model.

Finally, a comparison of the top and bottom panels of Fig. \ref{fig:f5_SiFe_vsFeH} shows that in our scheme the so-named High-Alpha Metal-Rich (HAMR) stars observed in the local disc can be attributed to intermediate-age stars (between  9 and 5 Gy) of the inner disc which migrated radially to the Solar neighborhood, as suggested already in \cite{Adibekyan2012}.

\subsection{Discussion on \afe \ vs \feh \  and the \feh \ gradient}
\label{subsec:afe_discussion}

The double sequence of \afe \ in the thin and thick discs of the Solar vicinity is fairly well established by now \citep[e.g.][and references therein]{Adibekyan2011,per21,Sharma2021b,Weinberg2022} although its extent varies a lot, depending on the chosen samples and biases: number of stars, volume limit of the sample, (mixture of) elements adopted as alpha, precision of measurements, etc. The difference in \afe \ for a given value of \feh \ between the two sequences may vary from less than 0.1 dex to more than 0.2 dex\footnote{In Sec. \ref{sub:X_Fe} we suggest a quantitative definition of that difference.}, the highest metallicity of the thick-disc sequence is not well constrained, and neither is the lowest metallicity of the thin disc.

The origin of this double sequence is not yet elucidated, and various suggestions have been made in the literature (see Sect. 1). In this work we confirm, as already argued in the study of \cite{Sharma2021} that secular evolution can produce naturally such a double sequence. In our model this results  as a consequence of:

a) inside-out disc formation, with a short timescale for the inner disc and a longer one for the outer disc.

b) ratio of two elements with sources evolving on widely different timescales: $\alpha$-elements from massive stars evolving in $\sim 10$ Ma, vs Fe from SNIa evolving in $\sim 1$ Gy timescale.

c) coexistence in the Solar cylinder of stars with appropriate histories of star formation, formed in the inner disc and locally.

The consequence of the above is that in the inner disc there is a large number density of stars in the \afe vs \feh \ plane that have large \afe \ values for low \feh, while in the Solar neighborhood the opposite happens. In intermediate regions, number density of stars {\it per unit area  of the \afe \ vs \feh \ plane} is smaller, as shown in Fig. \ref{fig:f5_SiFe_vsFeH}. Radial motions of stars through blurring and churning lead to point (c). In summary, the adopted radial profile for the infall rate in our model (see Fig. \ref{fig:f1_discs_gen}, top panel), is at the origin of the gap between the thin and thick disc in the \afe vs \feh \ plane. 

In their semi-analytical model, which has several different prescriptions than ours (in particular, parameterized chemical evolution)  \cite{Sharma2021} suggest a slightly different interpretation, attributing "the gap between the two sequences  to the sharp transition of \afe \- from a high value to a low value within a span of a few Gy, the transition being  due to time delay in the on-set of SNIa explosions."

In contrast, we note that the hybrid model of \cite{Johnson2021}, adopting several ingredients similar to our model and successful in reproducing several observables of the MW disc, fails to reproduce the observed bi-modality of \afe \ vs \feh \ in the Solar neighborhood. They suggest that a supplementary ingredient, like a two-phase star formation, would help in that respect.

At this point, we note that the recent study of \cite{Scott2021UGC10738} finds,  through stellar population
synthesis model fitting, that the galaxy UGC 10738 (a nearby, edge-on MW-like galaxy) contains alpha-rich and alpha-poor stellar populations with similar spatial
distributions to the same components of our Galaxy. They point out that if such features are found to be generic of MW-like galaxies, this may pose a challenge for the merger-triggered starburst theory of the origin of the Milky Way’s thick disc, because simulations suggest such events to be rare.  

This conclusion is in contrast with the recent work of \cite{Lu2022b} who analyze $\sim$80000 subgiant stars of LAMOST and find a non-monotonic of the \feh \  gradient  - taken in the birth places of stars -  around 8-11 Gy ago (see Sec. \ref{sub:Abundance_profiles} and Fig. \ref{fig:ModelvsAge_gradients2}). They associate it to the last major merger (Gaia Sausage/Enceladus event) and they  claim that this transition plays a major role in shaping the  [$\alpha$/Fe] vs [Fe/H] dichotomy. We show here and in Sec. \ref{sub:Abundance_profiles} that this may not necessarily be the case, and that secular evolution can also reproduce both those features.

\section{Other elementary ratios}
\label{sec:OtherXY}

In this section we explore the consequences of secular evolution for the ratios of other elements.

\subsection{[X/Fe] ratios}
\label{sub:X_Fe}

The case of \afe\- vs  \feh~ratio is a clear one because of the large difference between the timescales of massive stars and SNIa. However, the situation is less clear in other cases.
From the observational point of view it is important to identify to what extent a given element X displays a simple or a double sequence (thick vs thin disc) when plotted as [X/Fe] vs \feh. For that purpose, large samples of stars with precise spectroscopically defined abundances are required. 

For the present study, we adopted the AMBRE project data that
is described in \citet{lav13} and consist in the
automatic parametrisation of large sets of ESO high-resolution
archived spectra. Within the AMBRE project, the stellar atmosphere
parameters (in particular, T$_{\rm{eff}}$, log g, [Fe/H], and [X/Fe]
together with their associated errors) were derived using the
MATISSE algorithm \citep{rec06}. In particular the [X/Fe] ratios for the elements Mg, Mn, Ni, Cu and Zn were collected from the study by
\cite{mik17,san21}, those for S from \cite{per21} and, those for Ba, Eu, and Dy
from \citet{gui18}. All these studies are based on HARPS and/or FEROS  high resolution  and high signal-to-noise spectra within the AMBRE project. This ensures that the abundances adopted here were derived in a homogeneous way using identical spectroscopic analysis tools and method (atmosphere models, spectral line lists and spectral synthesis code) to derive the stellar parameters. We note that a similar analysis with the HARPS data and evaluation of the uncertainties per metallicity bin was performed by \cite{DelgadoM2017}, who also discussed some of the nucleosynthetic implications of the results.
%Furthermore, in order to increase the quality of the [X/Fe] ratios used to %compare with our model predictions, we selected only stars where the total %error estimated in the [X/Fe] ratios was lower than $\pm 0.1$ dex (Mg, S, %Mn, Ni, Cu, Zn, Ba) or $\pm 0.12$ dex (Eu, Dy) (see the studies above for %details). 
Then, we divided the selected stars according to their thin or thick disc membership following the criteria adopted by \citet{rec14} based on their [$\alpha$/Fe] ratio. The data for these 9 elements from AMBRE analysis are displayed in Fig. \ref{fig:X_Fe_vs_Fe_H}. For each element, the chemically separated thick and thin discs are shown with red and blue colours, respectively. 
%A typical error bar for the corresponding [X/Fe] and [Fe/H] ratios is also shown. 
%Boxes in the bottom of each panel contain the numbers of stars in each metallicity bin ($\Delta$[Fe/H]=0.2 dex). Curves join the average values of metallicity bins. 
%The grey portions of the curves correspond to metallicity bins shared by thin and thick discs (usually four bins). 
The common portions of the two curves are used to calculate {\it distances} in [X/Fe] between the thin and thick discs sequences, according to the {\it Z-test} \citep[e.g.][]{spr11}, as follows:

\begin{figure}
	\includegraphics[angle=0,width=0.49\textwidth]{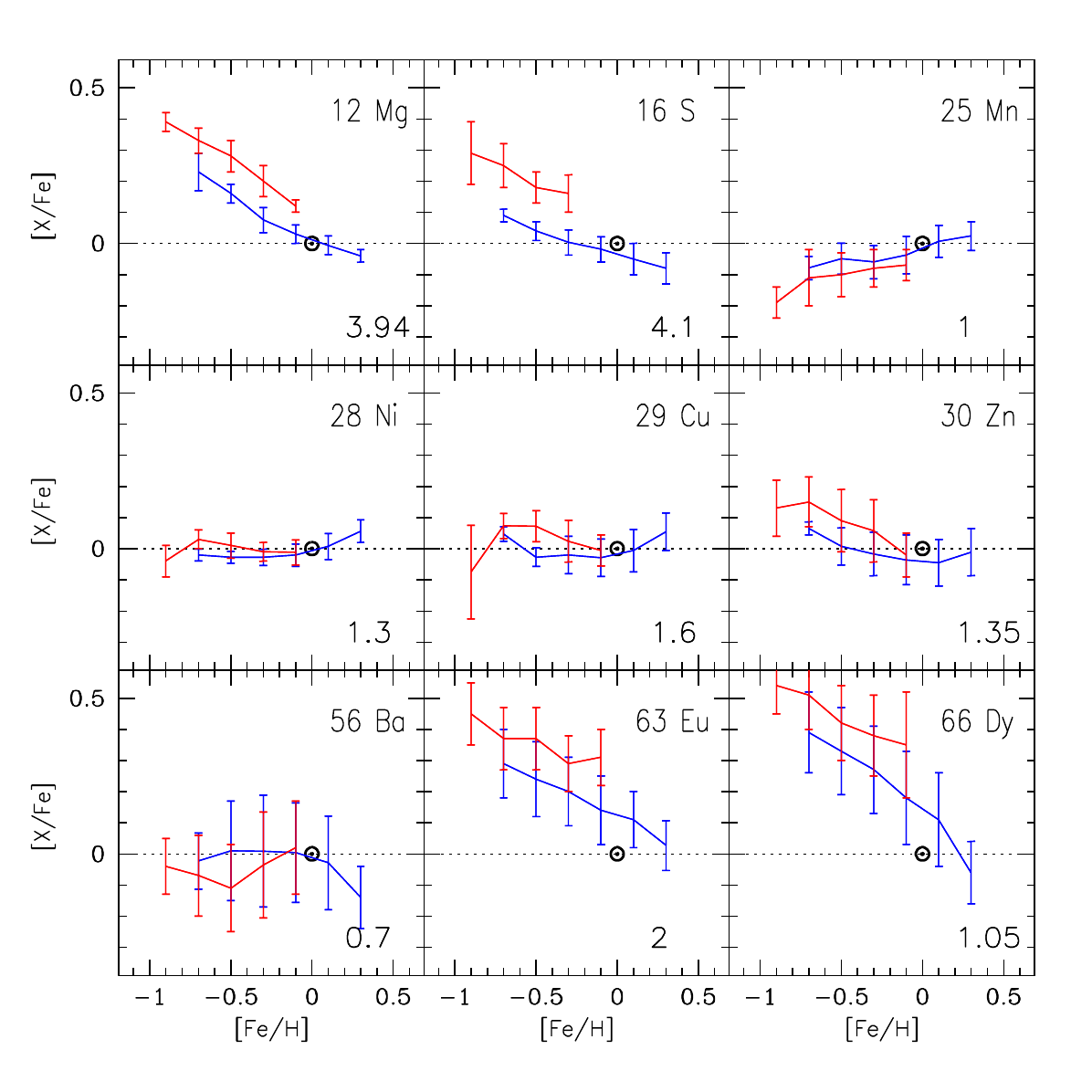}
    \caption{Observational data of [X/Fe] from the AMBRE project. Thin and thick disc stars are identified chemically (blue for thin and red for thick disc, respectively) on the basis of their \afe ~vs \feh ~behaviour (see text). 
    %Numbers within boxes indicate the number of stars in each metallicity bin. 
    Vertical bars are 1$\sigma$ uncertainties and curves connect the averages of metallicity bins. 
    %The grey portions of the curves correspond to metallicity ranges shared by the thin and the thick discs; they are used to calculate 
    The "distance" (D) between the two sequences is expressed by the numbers on the bottom right of each panel (see text for the calculation of the Z-test).
    }
    \label{fig:X_Fe_vs_Fe_H}
\end{figure}

At each \feh \ bin, the average values of [X/Fe] are determined,  $Y_T$ and $Y_t$  for the thick and thin discs, respectively, as well as the corresponding dispersion ($\sigma$).
%Assuming  the latter to be independent of the dispersion, the joint uncertainty is evaluated as  $u = \sqrt{\sigma^2+\epsilon^2}$.
 Then, the "distance" between the two distributions is calculated as D$={1\over N}\sum((Y_T - Y_t)/u)$, where N is the number of \feh \ bins which are common between the thin and thick disc stars (usually 4), and u is given by $u=\sqrt{\sigma_T^2+\sigma_t^2}$;  $\sigma_T$ and $\sigma_t$ are the standard deviation of the mean ($\sigma/\sqrt{m}$) in the corresponding metallicity bin. Here $m=$min$(n_T,n_t)$ where $n_T$ and $n_t$ are the number of stars in each metallicity bin for the thick and thin disc samples, respectively.
 
 Values of D larger than 3 indicate that the two sequences are sufficiently distant (more than one sigma) and therefore, different. On the contrary, values of D significantly lower than 2 indicate that the two data sets can not be distinguished. Values intermediate between 2 and 3 means that the samples are marginally different (D$< 2.5$), or significantly different ($2.5\leq D < 3$).
%taking into account the measured statistical uncertainties, or perhaps that those %uncertainties are underestimated. 
%On the other hand, values of D suggest either that the two sequences are really very close to %each other or that the associated uncertainties are still quite large. 
These numbers are indicated in Fig. \ref{fig:X_Fe_vs_Fe_H}. 
Mg and S display small dispersion in all their metallicity bins and the sequences of their mean values are clearly separated, so the corresponding overall Z-distances are high (around 4): observationally, these two elements have clearly different [X/Fe] sequences in the local thin and thick discs.   In other cases, (Mn and Ni), the sequences of mean values are very close and despite the small dispersion the corresponding Z-distances are close to 1, implying that these elements have a single sequence. For the remaining elements, low Z-distances are obtained, and this is due either to very close sequences of mean values (for Cu and Ba) or to large dispersion (for Zn and Dy). Eu appears to be a kind of intermediate case, with two clearly separated sequences (as for S), but the dispersion is rather large and leads to an intermediate value for the final Z-distance. Observationally, it is impossible with the available data to say whether Eu displays a double sequence or not.
%While Mg and S have small uncertainties and display clearly two distinct sequences for the thin and thick discs (D$>>1$), the situation is unclear for the Zn, Eu and Dy (D$\sim1$). In the cases of  Mn, Cu, Ni and Ba can not be distinguished two abundance sequences (D$< 1$). Note that the ambiguous result for Eu and Dy certainly is conditioned by the large average total error in [X/Fe]. A reduction of the uncertainty on [X/Fe] ratio would give a more clear picture on the existence of two sequences or not for these elements. Eu and Dy are mainly r-process elements which typically (REF) behave like alpha-elements in the [X/Fe] vs [Fe/H] relation at the metallicities studied here. Thus, a double sequence similar to that found for Mg and S might exist for both elements.

\begin{figure}
	\includegraphics[angle=0,width=0.49\textwidth]{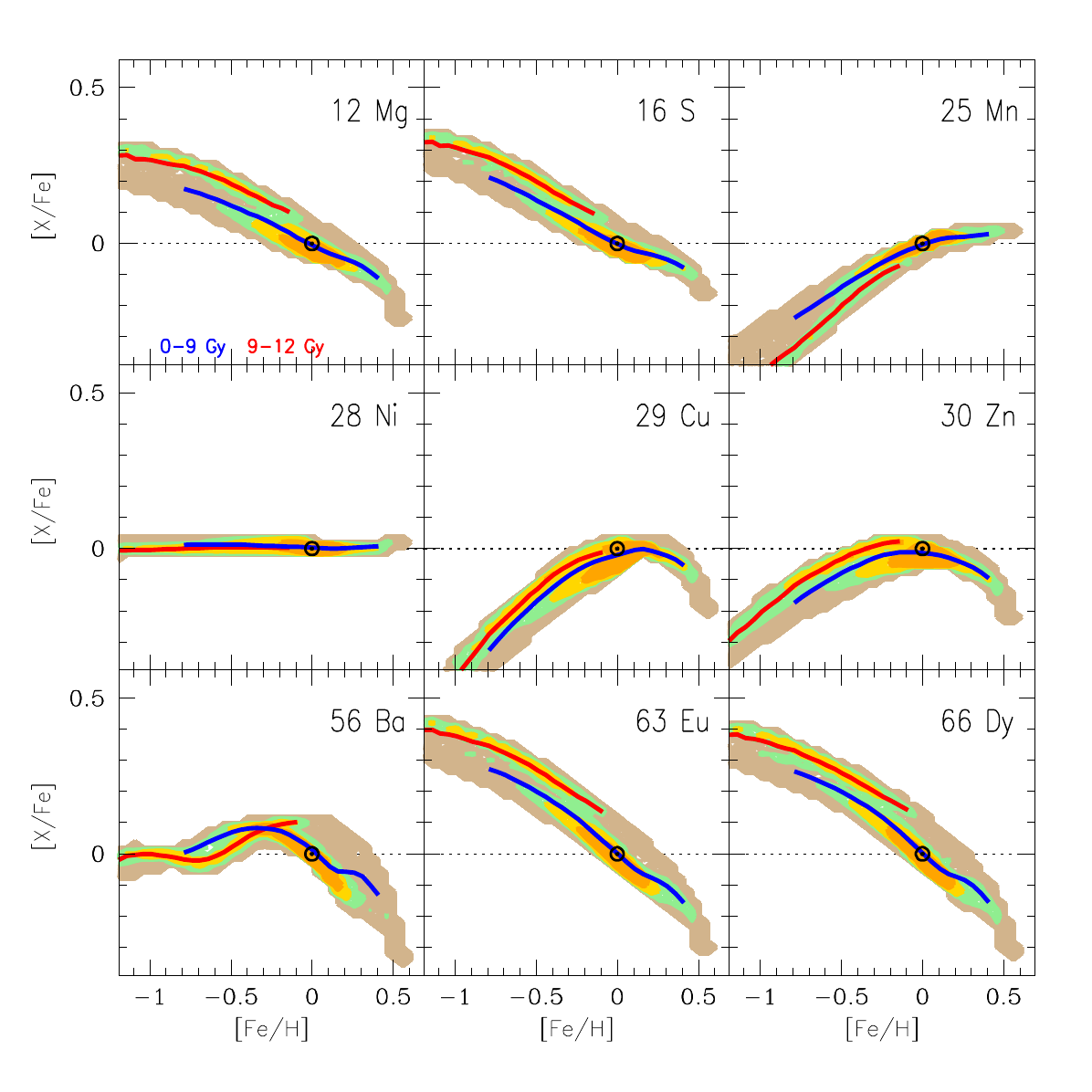}
    \caption{Model counterpart of Fig. \ref{fig:X_Fe_vs_Fe_H}. Colour coding as in Fig. \ref{fig:f5_SiFe_vsFeH}, with the red and blue curves indicating average values for the thick and thin discs, respectively: see text for discussion of individual elements.
    }
    \label{fig:X_Fe_vs_Fe_H_mod}
\end{figure}

In Fig. \ref{fig:X_Fe_vs_Fe_H_mod} we present our model results for the same elements. Colour coded isocontours separate clearly the thin from thick disc sequences in the cases of some elements: alpha elements, like Mg and Si (as discussed in Sec. 4.2), or r-elements, like Eu and Dy. The latter is due to the fact that we adopted here, for illustration purposes, collapsars (rotating massive stars) as the sources of r-elements, with their yields scaled to the ones of oxygen (see Eq. 2 in \cite{Prantzos2018}); the behaviour of those r-elements follows then naturally the one of oxygen in our model.

In contrast, there are cases where distinct sequences of abundance ratios [X/Fe] in the thin and thick discs do not appear.  The clearest one is the case of Ni, which behaves exactly as Fe, since both of them have the same sources: they are produced in the same proportions by nuclear statistical equilibrium in the explosions of CCSN (for about 30-50 \% of the corresponding Solar abundances) and SNIa (for the remaining 70-50 \%).

Manganese behaves as a pseudo-secondary element in the  thick disc in our model, in qualitative agreement with the observations, but with a more enhanced secondary-like behaviour than observed. This is due to the metallicity dependent yields of massive stars adopted here, which are also found in the case of the yields from \cite{Woo95} and \cite{Nomoto2013} (see discussion in \citet{Prantzos2018}). Because of the strong Fe production by SNIa, the [Mn/Fe] ratio of our model flattens in the late thin disc, in fairly good agreement with observations. It should be emphasized that the role of SNIa in the production of Mn is not yet completely understood (see discussions in \cite{Badenes2008,Kobayashi2020}). Our model suggests a very weak double sequence behaviour of [Mn/Fe] in the thick and  early thin discs, compatible with a single sequence. 

The [Cu/Fe] rises rapidly in the thick and thin discs of our model and displays a single sequence in Fig. \ref{fig:X_Fe_vs_Fe_H_mod}. The observations of Fig. \ref{fig:X_Fe_vs_Fe_H} show indeed a single sequence, but there is no significant rise of [Cu/Fe]. 
The rising of the Cu abundance  of the model is due to the secondary-like nature of the dominant isotope $^{63}$Cu, produced mainly by neutron captures in the
He-core for metallicities higher than [Fe/H]$\sim -2$ (see also discussion in \cite{Romano2007}.

In the case of Zn, our model fails to reproduce the observed trend of [Zn/Fe]$>0$ in the thick disc, as well as  in the halo \citep{Prantzos2018}. A supersolar ratio in the early Galaxy can be explained by invoking a considerable amount of hypernova  explosions \cite[]{Nomoto2013,Kobayashi2020}, which we did not introduce here. It would be interesting to see whether hypernovae could produce  a single or double-branch trend in the Galactic disc and whether their Zn yields depend as strongly on metallicity as the ones of massive stars adopted here. The massive star yields of Zn are mostly due to the weak s-process occurring hydrostatically in the He-core and this is the reason of their metallicity dependence, leading to a rising [Zn/Fe] at [Fe/H]$<-0.5$. At higher metallicities, the rising of Zn is somewhat compensated by the rising Fe from SNIa, leading to a flat profile of [Zn/Fe] with metallicity. We note that Zn is also produced in large amounts in the explosions of low-mass supernova ($\sim$8-10 M$_\odot$) through electron capture \citep{Wan13}; however, the impact of those sources is expected to be small, in view of their small number with a normal IMF.

The case of Ba is different. At low metallicities ([Fe/H]$<-1$) it is the r-component of Ba that matches the evolution of Fe produced by massive stars and the [Ba/Fe] ratio remains quasi-constant. Around [Fe/H]$\sim -1$, the efficiency of the main s-process from AGB is at its maximum \citep{Cr15}, but Fe production from SNIa starts becoming important and the [Ba/Fe] ratio barely increases in the late thick and early thin discs. At even higher metallicities (late thin disc), the efficiency of the main s-process is considerably reduced \citep{Cr15,Prantzos2018}, while Fe production from SNIa continuous and thus the [Ba/Fe] ratio slightly decreases. The result of the operation of all these sources (AGB with metallicity dependent Ba yields and lately appearing  SNIa with constant Fe yields) produce a quasi-constant [Ba/Fe] ratio over the whole history of the disc (except for the highest metallicities) and lead to a single-branch behaviour, as seen in both observations and model.

Summarizing the discussion of the models vs observations of this section we may say that for the [X/Fe] ratios in the thin and thick discs

- there are cases with double sequences (Mg and S) and single ones (Ni), which are sufficiently established observationally and well understood theoretically; Mn could be also put in that class, despite some uncertainties on the role of SNIa.

- the large dispersion in the abundance data does not allow yet to establish the existence of a single or double sequence for Ba, Eu or Dy observationally. 
Our models predict a single sequence for the s-element Ba and a double one for the r-elements Eu and Dy, but the latter result depends on the adopted assumption that r-elements are produced by collapsars (short-lived progenitors, as for the $\alpha$-elements).

- the observed approximately flat trend of [X/Fe] with [Fe/H] for Cu and Zn is not well reproduced by our models, but the uncertainty on the sources/yields of those elements and their behaviour with metallicity leaves a lot of room for improvement. In that respect, it would help to have a clear picture of whether a single- or double sequence exists for those elements. In the next subsection we made our ideas for the importance of such sequences more explicit. 

\subsection{[X/Y] ratios}
\label{sub:X_Y}

Trying to interpret the sequence(s) of [X/Fe] vs [Fe/H] in the previous subsection, we invoked two reasons related to the sources of the elements involved: 

i) the age of the system at the time of the enrichment of the ISM with the considered elements (because different ages allow for different types of sources to operate) ; and

ii) the initial metallicity of the sources, which affects in different ways the ejected amounts of the considered elements (primaries vs secondaries or pseudo-secondaries, or odd vs even elements).

One might conceive other causes affecting the evolution of a given element, e.g. a variable IMF (either in its slope or in its upper limit), but we stick here to the simplest case of two confirmed causes; thus we distinguish

- sources: short-lived (SL) vs long-lived (LL) ones; in the former case belong CCSN and collapsars  (characteristic time-scales $\sim$10 Ma) and in the latter AGB stars, SNIa and NSM (timescales 100 Ma to a few Gy)

- yields: metallicity independent (MI) for primary or even elements (alpha elements, Fe, r-elements?); and metallicity dependent (MD) for secondary elements (e.g. Ba) or odd elements (e.g. Na, Al).

\begin{figure*}
	\includegraphics[angle=270,width=0.99\textwidth]{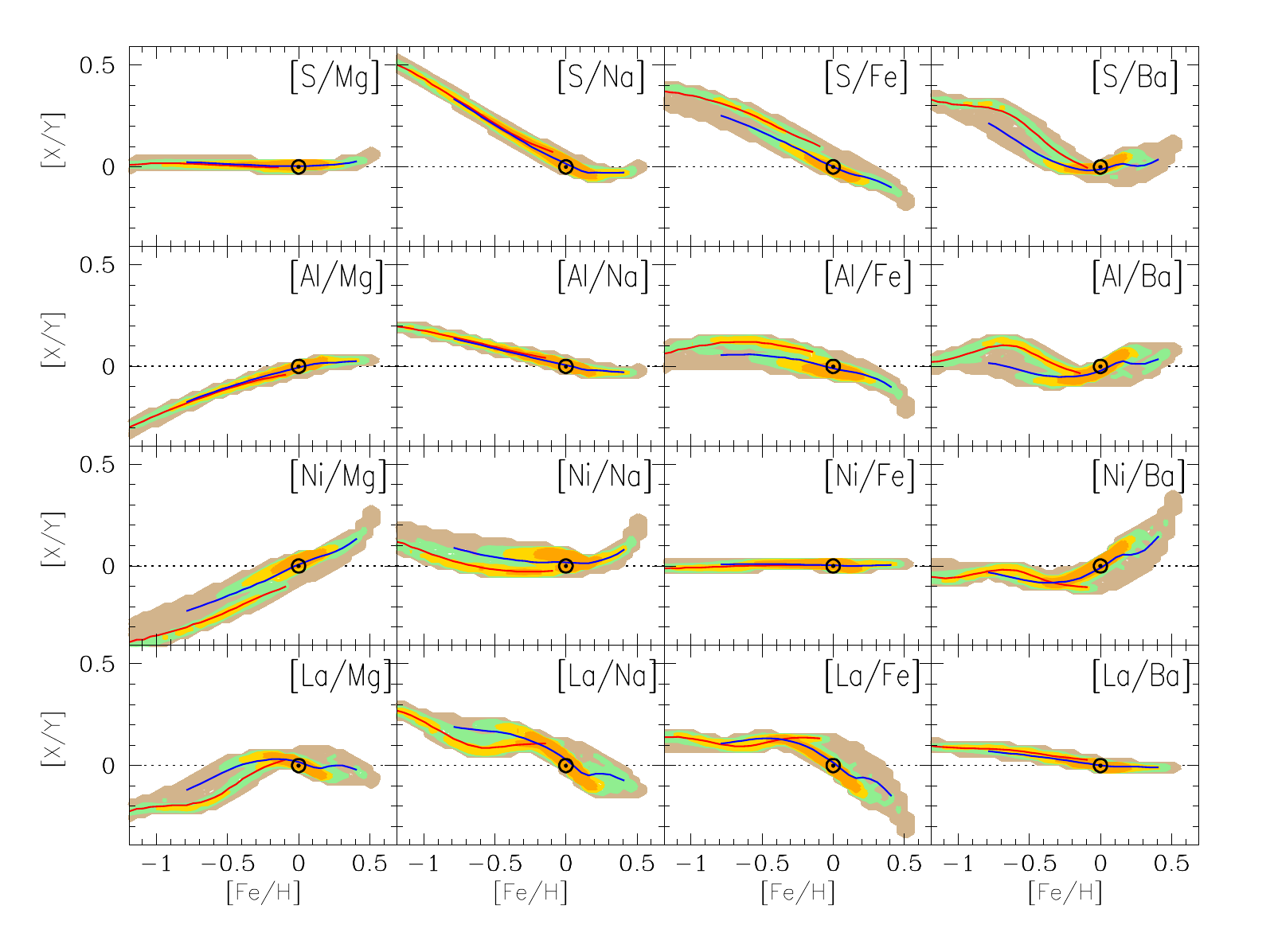}
    \caption{Model results for the Solar cylinder and for [X/Y] abundance ratios vs [Fe/H], for couples of elements X and Y arranged as in Table \ref{tab:Elements}.}
    \label{fig:X_Y_Localdisc}
\end{figure*}

This scheme provides four combinations for a given element, depending on its source and nature of its yield: a) SL-MI; b) SL-MD; c) LL-MI and d) LL-MD. For the ratio of two elements we get then 16 combinations, as illustrated in Table \ref{tab:Elements}. For each one of the cases (a) to (d), two elements (X, Y) belonging to that class are selected: (S, Mg), (Al, Na), (Ni, Fe) and (La, Ba), respectively; the first one serves in the nominator and the second one in the denominator of [X/Y]. The results of our simulations for [X/Y] vs [Fe/H] appear in Fig. \ref{fig:X_Y_Localdisc}. The simplest case is obviously when both X and Y have same sources and metallicity dependence of their yields. This is illustrated by the case of [S/Mg] (primary elements produced in CCSN) and [Ni/Fe] (primaries produced first in CSSN and later also in SNIa, at the same rate). The resulting sequences are the same (one branch or 1-B) and are flat (slope = 0).

\begin{table}
\begin{center}
\caption{ Expected 1-branch (1-B) or 2-branch (2-B) behaviour of abundance ratios of elements belonging to classes A, B, C and D (as defined below) in the local thick and thin discs. }
\label{tab:Elements}
\begin{tabular}{c c c c c c c}
\hline
%\\
  & A : SL-MI  &  B : SL-MD & C : LL-MI  & D: LL-MD\\
  &  {\it Mg} & {\it Na} & {\it Fe} & {\it Ba} \\
% \\
\hline

A : SL-MI &   1-B, s=0 &  1-B, s<0 &  2-B, s<0  &   2-B ? \\
 {\it S} &  {\it [S/Mg]} & {\it [S/Na]} & {\it [S/Fe]} & {\it [S/Ba]} \\
  \\
 B : SL-MD &   1-B, s>0 &  1-B, s$\sim$0 &  2-B, s$\sim$0  &  2-B ?  \\
 {\it Al} & {\it [Al/Mg]} & {\it [Al/Na]} & {\it [Al/Fe]} & {\it [Al/Ba]} \\
 \\
C : LL-MI &   2-B, s$>$0 &  2-B, s$\sim$0 &  2-B, s=0  & 1-B    \\
{\it Ni} & {\it [Ni/Mg]} & {\it [Ni/Na]} & {\it [Ni/Fe]} & {\it [Ni/Ba]} \\
\\
D : LL-MD &   2-B B &  ? &  1-B  & 1-B, s$\sim$0   \\
{\it La} & {\it [La/Mg]} & {\it [La/Na]} & {\it [La/Fe]} & {\it [La/Ba]} \\
\hline

\end{tabular}
\end{center}

{\it Lifetimes of nucleosynthesis sources }. SL: Short Lived ($\sim$10 My, CCSN); LL: Long Lived ($\sim$1 Gy, SNIa, AGB and perhaps NSM). \\
{\it Source nucleosynthesis yields}. MI: Metallicity Independent (primaries, even); MD: Metallicity Dependent (secondaries, odd) \\
s is the slope of the relation [X/Y] vs [Fe/H], with X on the left column and Y in the top raw.

\end{table}

In the same case (same sources and metallicity dependence of yields) belong the couples (Al, Na) produced by CCSN and (La, Ba) made mainly by AGBs. Here we also get a single sequence. However, the difference with the previous case is that the metallicity dependence of the yields obviously is not exactly the same for the two members of each couple. As a result, the single sequence has a slight metallicity dependence. Note that the position of the members of each couple is arbitrary: in the chosen cases, the slopes are negative, but by inverting the position of X and Y the slopes would be positive.

The next simplest case is when the two elements X and Y have the same source, but their yields have substantially different metallicity dependence (MD), like a primary vs a secondary (or an odd) element. Then a single sequence is also obtained, because of the same timescale of the evolution of sources, but it has a strong slope, either negative (when the MD element is in the denominator, like [S/Na]) or positive (when the MD element is in the nominator, as in the case of [Al/Mg]).

The next case is the one most extensively analysed, namely two primary elements with sources evolving on different timescales (SL and LL). This is typically the case of [$\alpha$/Fe] ratio, discussed in Sec. \ref{subsec:alphaFe_FeH}.  It leads to 2 quasi-parallel branches (2-B), as illustrated here for [S/Fe]: its slope is negative because the LL element (Fe) is in the denominator. The "mirror"-case is [Ni/Mg], with two sequences and  a positive slope, since the LL element (Ni) is in the nominator.

The previous cases involve elements having at least one common feature, either source lifetime or yield metallicity dependence, or both. The situation becomes more complicated when the two elements have no common feature. It is well known that the effects of metallicity dependence of the yields may mimic to some extent those of a source with long lifetime.

Those effects are illustrated in the case of [Al/Fe]. Al is a SL-MD element (see Table 1), while Fe is a LL-MI. In Fig. 
\ref{fig:X_Y_Localdisc} a 2-branch behaviour appears for the thin and thick discs, due to the difference in source lifetimes. On the other hand, the metallicity dependence of Al yields from CCSN compensates somewhat for the late production of Fe in SNIa and leads to a rather flat curve in the late thick disc and in the thin disc. The behaviour of [Ni/Na] is the mirror image of that, since the SL-MD element (Na) is now in the denominator. 

The s-elements Ba and La are displayed here only for illustration purposes, since their case is even more complex than the previous ones. They start with an r-component the importance of which is reduced at later times, because of the rising s-component from the main s-process in LL AGBs. This matches to some extent the rising contribution of Fe-peak elements from   SNIa and, along with the long lifetime of AGBs, contribute to produce a single branch behaviour. In the superSolar metallicity regime of the thin disc, the efficiency of the s-process is reduced and the ratio of those s-elements to Fe declines. These results are illustrated in the behaviour of [La/Fe] and its mirror image of [Ni/Ba].

Although our results on [La/Mg] and [La/Na] (as well as their mirror images of [S/Ba] and [Al/Ba]) seem to suggest a 2-branch behaviour, we refrain from any conclusions, since the corresponding over-densities in the [X/Fe] vs [Fe/H] plane for the early thin disc are fairly weak.

Finally, in the cases of [La/Fe] and [Ni/Ba], we observe that the late increase of the SNIa products (Fe and Ni), combined with the decrease of the efficiency of main s-process production (for Ba and La) at high metallicity, leads to a decrease of [La/Fe] and a corresponding increase of [Ni/Ba] for [Fe/H]$>0$. This result depends on both the adopted AGB yields at high metallicity and the rate of SNIa in the inner disc regions, where the super-Solar metallicity stars are assumed to be formed.

We remind that in the simulation studied here we assumed, for illustration purposes, production of the r-component of all heavy elements in collapsars evolving on short timescales. Those elements behave then obviously as alpha elements in our calculations, as already shown in Fig. 
\ref{fig:X_Fe_vs_Fe_H_mod} for Eu and Dy. However, since the source(s) of those elements still remain uncertain and they may evolve on a range a timescales, we prefer to postpone a more thorough discussion of that topic for a future paper.

It is interesting to notice that whenever a "2-branch" behaviour is obtained in our scheme, the two branches are reversed in the "mirror image": if the [X/Y] sequence of the thick disc is higher than  the one of  the thin disc for a given metallicity, it appears below it in the mirror image, as e.g. in the cases of [Ni/Mg] vs [Si/Fe] or [Ni/Na] vs [Al/Fe]. 

%\begin{figure}
%	\includegraphics[width=0.49\textwidth]{a4_new.pdf}
%    \caption{[$\alpha$/Fe] (left) and [$\alpha$/S] (right) vs [Fe/H] in the Solar neighborhood  with massive star yields from LC2018.  }
%    \label{fig:f5_alphaFe_vs_Age_LC18} \end{figure}

%\begin{figure}
%	\includegraphics[width=0.49\textwidth]{a5_new.pdf}
%    \caption{[Same as in Fig. \ref{fig:f5_alphaFe_vs_Age_LC18} but with massive star yields from Nomoto2012.}
%    \label{fig:f6_alphaFe_vs_Age_Nom06}
%\end{figure}

\begin{figure}
	\includegraphics[angle=270,width=0.5\textwidth]{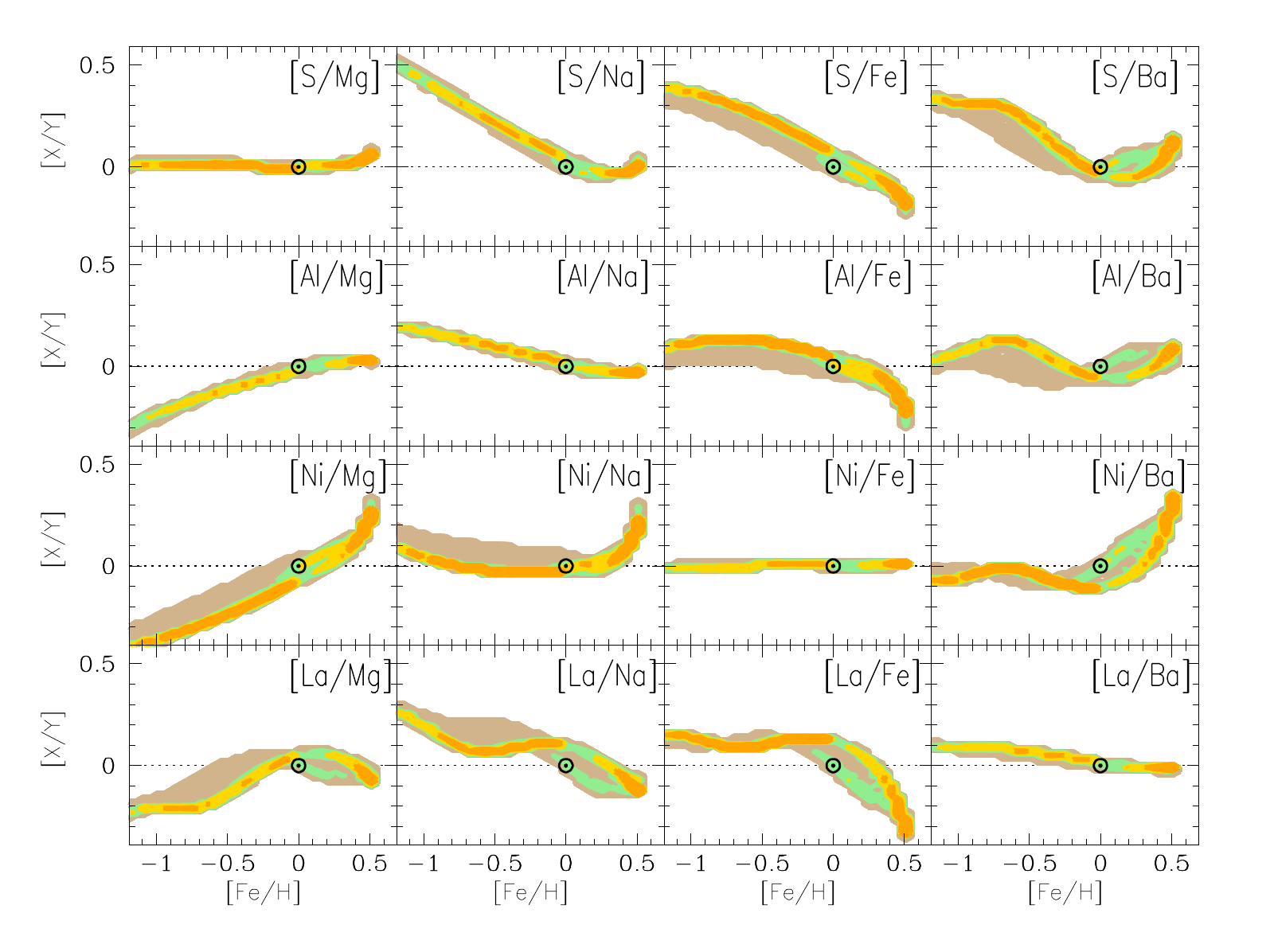}
    \caption{ Same as in Fig. \ref{fig:X_Y_Localdisc}, but this time for the inner disc, in the radial range of 2-3.5 kpc.}
    \label{fig:X_Y_Innerdisc}
\end{figure}

\begin{figure}
	\includegraphics[angle=270,width=0.5\textwidth]{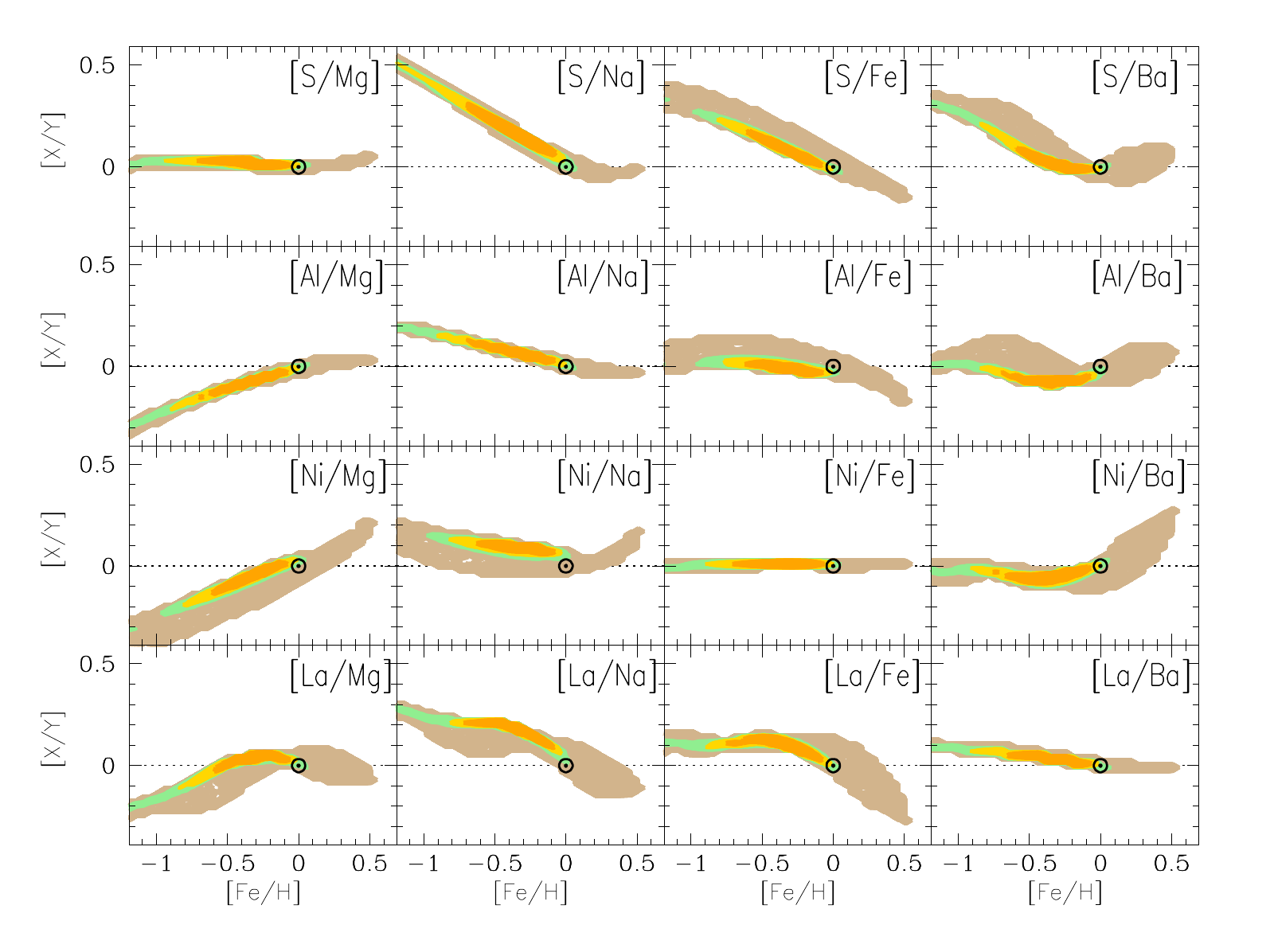}
    \caption{Same as in Fig. \ref{fig:X_Y_Localdisc}, but this time for the outer disc, in the radial range of 12-15 kpc. }
    \label{fig:X_Y_Outerdisc}
\end{figure}

The results displayed in Fig. \ref{fig:X_Y_Localdisc} and discussed in the previous paragraphs concern the local disc, and more precisely, the Solar cylinder (a region of width of $\Delta R \sim 1$ kpc center at Galactocentric distance of $\sim 8$ kpc). Results are different for other disc regions, as illustrated in Figs. \ref{fig:X_Y_Innerdisc} and \ref{fig:X_Y_Outerdisc}.

In Fig. \ref{fig:X_Y_Innerdisc} we show the results for the region 2-3.5 kpc. In this case the duality is very strongly diminished, so that one could say that there is a single sequence  (albeit with a finite width) for all element ratios. The reason is obviously that those inner regions evolved on similarly short timescales and thus have a common evolution, while they received little contribution from the outer disc which evolved on long timescales. A comparison with  Fig. \ref{fig:X_Y_Localdisc} reveals two more features. First, at subSolar metallicities, the unique sequence of Fig. \ref{fig:X_Y_Innerdisc} corresponds to the thick-disc sequence of the local disc (the red-coloured sequence in Fig.  \ref{fig:X_Y_Localdisc}). This is a clear signature that the local thick disc has a common origin with the inner disc. Secondly, there is an overdensity of stars at superSolar metallicities [Fe/H]$\sim +0.3-+0.5$ dex, which corresponds to the late saturation of metallicity in the inner disc (see the two bottom panels of Fig. \ref{fig:f1_discs_gen}). Such an overdensity does not exist in the local disc, but clearly its super-Solar metallicity stars come from those inner disc regions through radial migration, as discussed in Sec. \ref{sub:Local_AgeVsZ}.

In Fig. \ref{fig:X_Y_Outerdisc} we display the same results for the outer disc, in the region 12-14 kpc. Again, no double sequence is obtained for any element ratio. Those regions evolve on similarly long times scales and receive very little contribution from the inner disc through radial migration, thus they display again a single sequence in their elementary abundance ratios. In contrast to the previous case of the inner disc, this single sequence is close to the thin disc sequence of the Solar cylinder (the blue curves in Fig. \ref{fig:X_Y_Localdisc}). Furthermore, there are basically no super-Solar metallicity stars in those outer disc regions; very few of the large numbers of such stars formed in the inner disc manage to reach the outer disc through radial migration, since the required time to cross $\sim$10 kpc is $\sim$10 Gy.

We note that a direct comparison of our results with observations is not always feasible, since our model is a 1D  and provides results for a given Galactocentric radius, but not as a function of distance from the plane. On the other hand, because of dust extinction in the Galactic plane, existing observations provide results for the inner and outer discs mostly  at high galactic latitudes , which are representative essentially of the thick disc.  For instance \cite{Lian2020MNRAS} select $\sim$40,000 stars from  Sloan Digital Sky
Survey (SDSS) DR16 with high signal to noise ratio and chose to define the thick disc geometrically, as stars at distances d$>1$ kpc from the mid-plane of the Galactic disc. When  the ensemble of all stars in the region $4<$ R (kpc) $<8$ is considered (left panel in  their Fig. 2), two overdensities appear in the [$\alpha$/Fe] vs [Fe/H] diagram, in the top left (high [$\alpha$/Fe], low [Fe/H]) and bottom right (low [$\alpha$/Fe] and super-Solar [Fe/H]). Both those features appear also in our Fig. \ref{fig:X_Y_Innerdisc} for the inner disc. They are due to the fact that at late times the abundances of most zones in the inner disc reach a "saturation equilibrium", where the metals produced by the few stars newly formed are diluted by the metal-poor gas ejected by the numerous stars formed in  these zones during the early times; for that reason, the number density of stars is high in that region of the \afe \ vs \feh \ plane.

\subsection{Discussion on [X/Y] ratios}
\label{subsec:OtherXY-discusssion}

In the previous section we attempted a classification of the various abundance patterns that are expected to arise in the thin and thick discs of the Galaxy, on the basis of the properties of the corresponding nucleosynthesis sources and yields. Indeed, in view of the large number of existing and forthcoming data, some classification scheme is required for the expectations to be compared to observations. We discuss here two recent attempts in that direction.

\cite{Weinberg2019} used $\sim$20,000 stars of the upper red giant branch from APOGEE DR14 survey to map the trends of elemental abundance ratios across the Galactic disc, spanning  the radial range $R= 3-15$ kpc and mid-plane distance $|z|=0-2$ kpc for 15 elements: O, Na, Mg, Al, Si, P, S, K, Ca, V, Cr, Mn, Co, and Ni. In view of the uncertainties on Fe production and evolution (due to mass cut and explosion energy in SNII, or to the uncertain evolution of the rate of SNIa), instead of using Fe as a reference element (i.e. [Fe/H] as a proxy for time) they chose Mg, which is well measured by APOGEE and is solely produced by CCSN \footnote{For similar reasons \cite{Goswami2000}  suggested to replace Fe  by an alpha-element like O or Ca (see their Sec. 5.7 and Fig. 8)}.
Separating the stars in two populations according to their [Fe/Mg] ratio, the work of \cite{Weinberg2019} revealed that the median trends in the [X/Mg] vs [Mg/H] plane in each population are almost independent of location in the Galaxy, and that the observed trends can be explained with a semi-empirical "two-process" model that describes both the ratio of CCSN and Type Ia  contributions to each element and the possible metallicity dependence of the supernova yields. They conclude that these observationally inferred trends can provide 
strong tests of supernova nucleosynthesis calculations, at least in the simple framework they consider (i.e. neglecting variations of the IMF or the CCSN range with time or metallicity). %and that the features observed in the metallicity distribution functions (number of stars vs. [Mg/H]) as a function of the galactocentrinc distance and $|z|$) can be understood in terms of the stellar migration  and "upside-down" formation of the Galactic disc. 
They do not attempt, however, to characterize  those observed patterns in terms of the existence of single or double abundance sequences.

More recently, \citet{SharmaHayden2022} studied the dependence of elemental abundances on stellar age and metallicity among a sample of about 50,000 main-sequence turn off (MSTO) and $\sim 3,700$ giants Galactic disc stars from the GALAH DR3 survey \citep{Buder2021}. Ages and distances of the stars were computed from observed parameters by making use of stellar isochrones and parallax in the case of MSTO stars, or asteroseismic observables in the case of giant stars \citep[see][for details]{sha18}. These authors found that elemental abundances for 23 elements show trends with both age and metallicity. They studied the relation between the variation of [X/Fe] with metallicity and age separately. The relation and the relationship is well described by a simple model in which the dependence of the [X/Fe] ratio on age and [Fe/H] are additively separable. Then, elements can be grouped in the age-[Fe/H] plane and the different groups can be associated with different nucleosynthetic sites: massive stars, exploding white dwarfs, and AGB stars. Since the stars studied cover a large range in Galactocentrinc $(R,|z|)$ locations, and the trends found are similar to both MSTO and giant stars, the authors suggested that their results may be valid over the whole disc of the Galaxy. This result may have significant implications for Galactic archaeology since it makes possible to estimate the age and birth radius of the stars using abundances, and from this dynamical processes like radial migration  and star formation histories at different radial zones can be explored.

The project of \cite{SharmaHayden2022} is attractive, because it is based on simple assumptions: the composition of a  star is uniquely characterized by its formation  time and radial position and that the interstellar medium is fully homogeneous (chemically) at any time in a given place. This may indeed apply, at least to a first order, as discussed in \cite{SharmaHayden2022} and  suggested by observational arguments and semi-analytical models, which show that the trajectories of various radial zones in the planes of age-metallicity or [X/Fe]-metallicity are monotonic and do not mix with each other \citep[e.g.][and this work]{Schon2009,Kubryk2015a,Johnson2021,Sharma2021}. However, it may not apply in cases where the star formation did not evolve smoothly, e.g. through episodes of intense star formation like those recently reported for the Galactic disc \citep[][]{Mor_2019,RuizLara_2020,Sahlholdt_2022}: depending on their intensity and spatial extension, such episodes may lead to a mixing of the aforementioned trajectories and to degeneracy in determining the time and place of birth of a star from its  composition. Similar effects - albeit less important - could result from localised intense episodes  of infall. 

Independently of the aforementioned complications, \cite{SharmaHayden2022} use the slopes of the
observed [X/Fe] vs age  or metallicity relations for rather large ranges of those parameters (3 to 11 Gy for ages and -0.6 to 0.2 in \feh). The derived unique slopes are then used to group the various elements X in the plane of $\rm{\Delta[X/Fe]/\Delta\feh}$
vs $\rm{\Delta[X/Fe]/\Delta(log(Age)}$. However, the large ranges of the adopted parameters may mask important variations of the adopted slopes and lead to spurious results in the classification of the elements in that plane. For instance, an inspection of Fig. 9 in \cite{SharmaHayden2022} shows that Na appears to be the element with the closest behaviour to Fe, much closer than Ni, while on the other hand Ba and La appear to behave in quite different ways, although they  both belong to the second peak of the s-process.

Our approach is less ambitious than the one of either \cite{Weinberg2019} or \cite{SharmaHayden2022}, because it does not aim to provide a quantitative classification of the elements in terms of their nucleosynthetic origin, but only a qualitative one: the existence of one or two branches in [X/Fe] (or [X/Y])  vs \feh \ relation.

%\begin{figure*}
%	\includegraphics[angle=270,width=0.7\textwidth]{a7_new.pdf}
%    \caption{{\bf OK UP TO HERE, END OF THEORY}}
%    \label{fig:MgFe_FeH_t1_t2_t3}
%\end{figure*}

\section{Summary and conclusions}
\label{sec:Summary}

In this study we present an updated version of the 1D semi-analytical model of \cite{Kubryk2015a,Kubryk2015b}, describing the evolution of the Milky way disc. The model includes a parameterized treatment of radial migration inspired by N-body+SPH simulations, which accounts both for "passive tracers" of chemical evolution and "active long-lived agents" of nucleosynthesis (like SNIa and AGB stars). 

The model reproduces fairly well
several observed features of the present day MW disc, like the recently observed gradients of \feh \ and \afe \ from Open Clusters \citep[][Gaia-DR3 results]{Recio-Blanco2022b} or the up-turn in the local age-metallicity relation for super-Solar metallicity stars, anticipated in \cite{Kubryk2015a} and observed by \cite{Feuillet2018,Feuillet2019}, which it believed to constitute a clear signature of radial migration in the thin disc (see Sec. \ref{sec:Model_discs}).

A key feature of the model is the formation of the thick disc by secular evolution. In our case this occurs  as a
consequence of (Sec. \ref{sec:alpha_Fe}):

a) inside-out disc formation, with a short timescale for the inner disc and a longer one for the outer disc.

b) ratio of two elements with sources evolving on widely different
timescales: $\alpha$-elements from massive stars evolving in $\sim$ 10 My, vs
Fe from SNIa evolving in $\sim$1 Gy timescale.

c) coexistence in the Solar cylinder of stars with appropriate histories of star formation, the thick disc being made from stars formed early on in the rapidly evolved inner disc.

This scenario for the formation of the thick disc can explain its chemical properties, like  the long established double-branch behaviour of \afe \ vs \feh \ in the Solar neighborhood. It can also explain the recently evaluated non-monotonic evolution with age of the metallicity gradient, as inferred by \citep{Lu2022b} for the birth radii of stars currently present in the Solar neighborhood: after a steady decrease, the gradient becomes flatter at the oldest ages, a behaviour remotely reminiscent of the one concerning not the birth radii, but the current Galactocentric radii of stars. In contrast to \cite{Lu2022b} who favor early mergers as an explanation of that behavior, we argue here that it may result from a combination of radial migration {\it and} a variable (non-unique) slope of the abundance profiles across the Galactic discs (Sec. \ref{sub:Abundance_profiles}).

However, these successes do not  imply that  secular evolution was the only factor shaping the thick and thin discs.  The high velocity dispersion of old stars  can be attributed to an early highly turbulent phase - as assumed here - but also to early mergers, as advocated in various studies (see Sec. \ref{sec:Intro} for references). Our scenario, in line with e.g. \cite{Schon2009,Loebman2011,Hayden2017,Sharma2021,SharmaHayden2022}, does not require some particular event (a merger, or the paucity of star formation on Gy timescales and/or intense episodes of infall) to explain the chemical properties of the thick disc.

In Sec. \ref{sec:OtherXY} we explore the consequences of our model for other abundance ratios [X/Y] in the MW disc.   We suggest that a key parameter in such studies is the "abundance distance" separating the thick from the thin disc sequences of [X/Y]. We illustrate that suggestion by adopting a simple $Z-test$ to measure that distance in the case of a set of high resolution observations analyzed in the AMBRE project \citep{lav13} for a handful of elements (Mg, S, Mn, Ni Cu, Zn, Ba, Eu and Dy). We show that, barring observational uncertainties, that technique has a high diagnostic potential, since it separates clearly elements with sources evolving on different timescales, like e.g. $\alpha$-elements from Fe. In that respect, it would be interesting to have a precise evaluation of the abundance distance between the thin and thick discs for r-elements, their main source being poorly known at present. If collapsars are the main source, then a double branch would be expected, as displayed in Fig. \ref{fig:X_Fe_vs_Fe_H} for [Eu/Fe]; in contrast, if the main source is neutron star mergers, the thick and thin disc branches would be expected to be very close. 

Finally, we generalize that method to the ratios of any two elements X and Y, with nucleosynthesis sources differing either in one or both of their two key properties, namely timescale of their evolution and metallicity dependence of their yields. We find that it is the timescale difference that plays always the key role in producing distinct [X/Y] sequences for the thin and the thick discs, while the metallicity dependence of the corresponding yields plays rather a minor role.

We emphasize that the above conclusions/predictions  are obtained within the adopted framework of the formation of the thick disc. Other modes of thick disc  formation, e.g. through the action of early galaxy mergers, may lead to different chemical signatures and should be  investigated as well.

\section*{Acknowledgements}
We are grateful to the referee for a fairly constructive report.
EA acknowledges support from the Centre National d'Etudes Spatiales (CNES), France.
This work has made use 
of data from the European Space Agency (ESA) mission Gaia (https://www. 
cosmos.esa.int/gaia), processed by the Gaia Data Processing and Analysis Consortium (DPAC, https://www.cosmos.esa.int/web/gaia/dpac/ 
consortium). Support of ESO, OCA and CNES is acknowledged for the AMBRE project. Some of the calculations have been performed with
the high-performance computing facility SIGAMM, hosted by OCA. C.A. acknowledges partial support by project PGC2018-095317-B-C21 financed by the MCIN/AEI FEDER “Una manera de hacer Europa”, and by 
project PID2021-123110NB-I00 financed by MCIN/AEI
/10.13039/501100011033/FEDER, UE.

\section*{Data availability}
Data sets generated during the current study are available from the corresponding author on reasonable request.

%%%%%%%%%%%%%%%%%%%%%%%%%%%%%%%%%%%%%%%%%%%%%%%%%%



\bibliographystyle{mnras}
\bibliography{Prantzos} % if your bibtex file is called example.bib

\bsp	% typesetting comment
\label{lastpage}
\end{document}